\documentstyle[aps,prl,multicol,epsf]{revtex}
\global\firstfigfalse
\global\firsttabfalse
\renewcommand{\narrowtext}{\begin{multicols}{2} \global\columnwidth20.5pc}
\renewcommand{\widetext}{\end{multicols} \global\columnwidth42.5pc}
\input{epsf.tex}

\def\top#1{\vskip #1\begin{picture}(290,80)(80,500)\thinlines \put(
65,500){\line( 1, 0){255}}\put(320,500){\line( 0, 1){
5}}\end{picture}}
\def\bottom#1{\vskip #1\begin{picture}(290,80)(80,500)\thinlines \put(
330,500){\line( 1, 0){255}}\put(330,500){\line( 0, -1){
5}}\end{picture}}

\def\Tr{\mbox{Tr}}

\begin{document}

\title{Electron--electron Interactions in Disordered Metals: 
Keldysh Formalism}

\draft

\author {Alex Kamenev,$^a$ and Anton Andreev,$^b$ }

\address{
$^{a}$Department of Physics 
University of California Santa Barbara, CA  93106-4030, U.S.A.\\
$^{b}$ University of Colorado, Department of Physics, CB 390, 
Boulder, CO 80309-0390 U.S.A.\\
{}~{\rm (\today)}~ \medskip \\ \parbox{14cm} 
{\rm
We develop a field theory formalism for the disordered interacting
electron liquid in the dynamical Keldysh formulation. This formalism
is an alternative  to the previously used replica technique. In
addition it naturally allows for the treatment of non--equilibrium
effects. Employing the gauge invariance of the  theory 
and carefully choosing  the saddle point in the $Q$--matrix manifold, 
we separate  purely phase effects of the fluctuating 
potential  from the ones that change quasi--particle dynamics. 
As a result, the cancellation of super--divergent 
diagrams   (double logarithms in $d=2$) is automatically build in the 
formalism. As a byproduct we derive a non--perturbative expression for
the single particle density of states. The remaining low--energy
$\sigma$--model describes the quantum fluctuations of the electron
distribution function. Its saddle point equation appears to be the
quantum kinetic equation with an appropriate collision integral along
with collisionless terms. Altshuler--Aronov corrections  
to conductivity are shown to arise from the one--loop quantum
fluctuation effects.  
 }\bigskip \\ }

\maketitle

\narrowtext

\section{Introduction.} 
\label{sec:intro}

The physics of weakly disordered interacting electron systems at low
temperatures has been a subject of considerable theoretical and 
experimental interest over the past years (for review see Refs. 
\cite{Efros,Lee85} ).
Although significant progress has been made in
this direction, many thermodynamic and transport 
properties of such systems are not completely  understood and continue 
to stimulate both experimental and theoretical research. 
The latest revival of interest to the problem was prompted by the
 experimental discovery of a possible metal--insulator transition 
originally in clean Si  MOSFETS \cite{Kravchenko} 
and later in  and p-type GaAs \cite{Shahar}.

The low temperature behavior of the conductivity of a metal is mainly
determined by the quantum (weak localization) \cite{Gorkov79,Bergmann84} 
and the interaction 
\cite{Altshuler79} corrections to the classical Drude result. These corrections
are especially strong in low dimensional ($d\leq 2$) systems.
In two-dimensions, for example, both the lowest order weak
localization correction \cite{Gorkov79} and the lowest order interaction 
correction \cite{Altshuler79} diverge logarithmically at low temperatures.
The ultimate faith of the low temperature phase is
determined by the interplay between them.

According to the scaling theory of localization \cite{gang4}, in the
absence of electron-electron (e--e) interactions (and with no 
spin-orbit scattering), the quantum corrections lead to localization
of all single particle states in dimensions $ d \leq 2$ and thus to
insulating behavior for arbitrarily weak disorder (weak localization).  
Wegner \cite{Wegner79} proposed a replicated $\sigma$-model to study this 
problem. With the coupling constant corresponding to 
the dimensionless conductance $g$, this $\sigma$-model provided 
justification for the one-parameter scaling theory of 
localization \cite{gang4}. Later, Efetov \cite{Efetov82} introduced the 
supersymmetric version of the $\sigma$-model which obviated the 
need to take the tricky \cite{Verbaarschot86} zero  replica number limit.

Finkel'stein \cite{Finkelstein83} developed a replicated $\sigma$-model
approach for interacting disordered electron systems, 
which was farther developed in \cite{Castellani86,Belitz94}. He 
demonstrated further its renormalizability in the one-loop approximation 
and obtained the one-loop renormalization group  flow
equations. From these equations it followed that the weak coupling
fixed point corresponding to non-interacting metal is unstable.
The need for introducing  replicas in Finkel'stein's approach 
follows from the fact that the ensemble averaged
observables are obtained as derivatives of the {\em averaged 
logarithm} of the partition function. The formalism in
Ref.~\cite{Finkelstein83,Castellani86,Belitz94} 
utilizes Matsubara representation and is
therefore restricted to the equilibrium situation. 

Later it was suggested that the Keldysh type field theory, 
originally developed for the treatment of non--equilibrium systems 
\cite{Keldysh} may be an alternative to the replica  technique  
\cite{Aronov,Lu,Babichenko}. The point is that the use of the 
Keldysh closed contour in the time direction, leads to an automatically 
normalized (disorder independent !) partition function. This
circumvents the need to introduce replicas. 
A similar situation exists in the theory of spin glasses, 
where in addition to the replica approach \cite{Edwards}
the Martin-Siggia-Rose formalism \cite{MSR}, analogous to the Keldysh 
approach \cite{MSR,Sompolinsky,Feigelman}, has been used. 
This formalism provided insight complementary to that gained from 
the replica approach. Horbach and Sch\"{o}n \cite{Shon} 
developed a $\sigma$-model  for  non--interacting electrons in the Keldysh 
formalism.  Although  our treatment differs from their in  many important 
details, we have benefited much from their work.

Here we apply the Keldysh formalism to disordered interacting systems. 
We restrict ourselves to the consideration of spinless electrons in the
presence of a weak magnetic field (unitary ensemble) and leave the
considerations of the spin and Cooper channels for  future work.
Another important distinction of the present theory from the 
previous ones \cite{Finkelstein83,Castellani86,Belitz94} 
is the different choice of a saddle 
point of the functional integral on the $Q$--matrix  manifold. 
 The saddle point in
our formalism explicitly depends on a fluctuating potential in
the system (the Hubbard-Stratonovich field, which decouples the
electron-electron interaction). This choice of the saddle point allows
us to separate  pure phase effects of the fluctuating potential 
and to present the first, in our opinion, clear derivation of
the tunneling density of states (DOS) in a metal film 
obtained earlier by Finkel'stein \cite{Finkelstein83} and Levitov and
Shytov \cite{Levitov} by different means. 
Another advantage of this choice of the saddle point
is that the perturbative expressions for gauge invariant quantities
contain only single logarithms of temperature or frequency (in $d=2$). 
The diagrams containing double logarithms which appear in the
standard diagrammatic expansion \cite{Altshuler80} or in Finkel'stein's
formalism \cite{Finkelstein83} and cancel each other for any gauge
invariant quantity do not appear in our formulation at all.  This
significantly reduces the number of diagrams in each order of
the perturbation theory. We then obtain  a low energy theory in the
form of a   
$\sigma$--model. The advantage of the Keldysh formulation is that it allows 
for a clear physical interpretation of the effective degrees of freedom.  
They turn out to be the quantum fluctuations of the electron distribution 
function. The saddle--point equation on the massless manifold is just 
the quantum kinetic equation with an appropriate collision integral. 
The one--loop fluctuations on top of this saddle point lead to
corrections to various observables and in the case of conductivity 
can be identified with the Altshuler--Aronov corrections 
\cite{Altshuler79,Altshuler85}.

The paper is organized as follows: In section \ref{sec2} we present 
the functional integral representation of the Keldysh partition function 
for disordered interacting electron systems. Section \ref{sec.smodel}
is devoted  to the choice of an interaction--dependent saddle point
and the derivation of  an effective $\sigma$--model as the massless
fluctuations around this saddle  point. We discuss some applications
of the theory, like the derivation of the non--perturbative expression for
the single--particle Green function, in  section
\ref{sec.applic}. Quantum fluctuations and Altshuler--Aronov
corrections to the conductivity are the subject of section
\ref{sec.fluct}. In section \ref{sec.kinetic} we  derive the quantum
kinetic equation as the saddle point equation on the massless
manifold and discuss the corresponding collision relaxation times
along with the collisionless terms. Finally, in section
\ref{sec.discussion} we briefly  discuss the obtained results and the
future perspectives.

\section{Functional Integral Formulation}
\label{sec2}

\subsection{Keldysh formalism}
\label{subsec.kel}

Consider a unitary evolution of a system along a closed contour ${\cal C}$ in 
the time direction which consists of the propagation 
from $t=-\infty$ to $t=+\infty$ 
and then back from $t=+\infty$ to $t=-\infty$. All external time--dependent 
fields are assumed to be {\em exactly} the same during the forward and 
backward evolution processes. As a result, at the end of such evolution 
the system must find itself precisely in the original state. We thus 
conclude that the evolution operator
\begin{equation} 
\hat U_{\cal C}\equiv 1\, .
                                                              \label{e0}
\end{equation}    
Let us consider next the partition function defined as 
\begin{equation} 
Z=\Tr\{\rho_0 \hat U_{\cal C}\} / \Tr\{\rho_0 \} =1\, . 
                                                              \label{e1}
\end{equation}
where $\rho_0$ is a density matrix of the system at the initial 
time, $t=-\infty$, before the interactions and disorder are 
adiabatically switched on. A more informative object is 
the generating functional, which is obtained by 
introducing source fields. It is clear that to have a generating
functional not identically equal to unity, the source fields should 
have a different behavior 
on the forward and the backward parts of the contour. 
To shorten the subsequent expressions 
we shall operate with the partition function,
Eq. (\ref{e1}), and will introduce the 
generating functional in section \ref{subsec.sources}.

\begin{figure}
\vglue 0cm
\hspace{0.01\hsize}
\epsfxsize=0.9\hsize
\epsffile{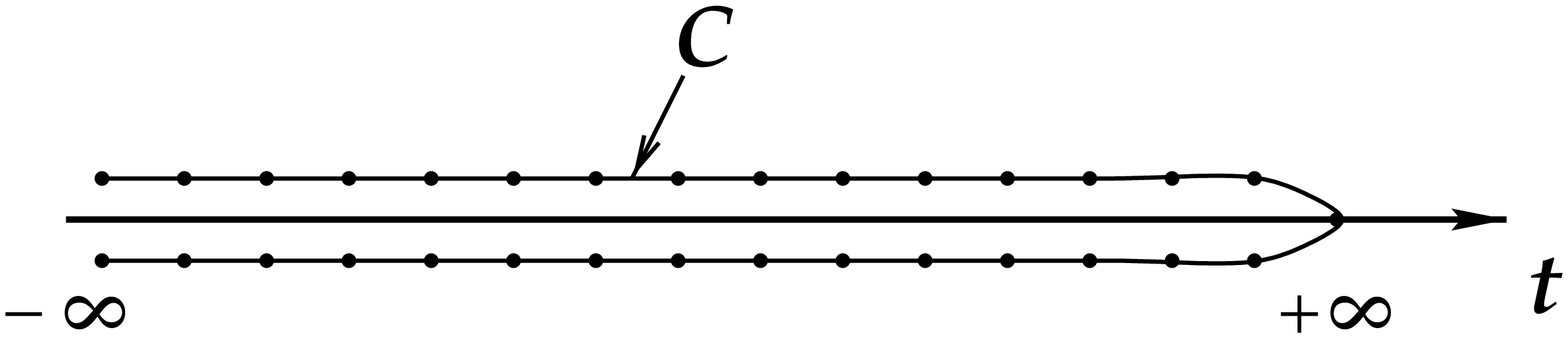}
\refstepcounter{figure} \label{contour}
{\small FIG.\ \ref{contour}
Schematic representation of the discretization of the time contour 
${\cal C}$. The dots on the upper and the lower branches of the contour 
denote the discretized time points.
  \par}
\end{figure}

The next step is to divide the ${\cal C}$ contour into $2N+1$ time
steps, such as $t_1=t_{2N+1}=-\infty$ and $t_{N+1}=+\infty$ as shown in
Fig.\ref{contour}.  Following the
standard route \cite{Negele}, we obtain the coherent state functional
integral, by introducing a resolution of unity at each time
step. Taking the $N\to \infty$ limit we obtain for
the partition function  
\begin{equation} 
Z={\cal N} \int {\cal D}\overline\psi \psi \exp\{iS[\overline\psi, \psi]\}\, ,
                                                              \label{e2}
\end{equation}
where ${\cal N}$ is {\em disorder independent} normalization
constant \cite{foot1}  and the fermionic action is given by
\begin{eqnarray} 
S[\overline\psi, \psi]= && \int\limits_{\cal C}dt 
\Bigg\{
\int d{\bf r} \overline\psi [G_0^{-1} - U_{dis}({\bf r})] \psi - 
                                                           \label{e3} \\
\frac{1}{2} &&\int\!\!\!\int d{\bf r} d{\bf r}' \overline\psi({\bf
r})\overline\psi({\bf r}')  
V_0({\bf r}-{\bf r}')\psi({\bf r}')\psi({\bf r}) 
\Bigg\}.
                                                              \nonumber
\end{eqnarray}
Here the inverse bare Green function is a shorthand notation for 
\begin{equation} 
G_0^{-1} = i\frac{\partial}{\partial t} + \frac{\nabla_r^2}{2m}   \, ,
                                                              \label{e4}
\end{equation}
where the time derivative is taken along the contour ${\cal C}$. 
The notation (\ref{e4}) is somewhat symbolic:  while inverting this
operator it is necessary to invert its discretized version first, and only
then take the limit $N\to\infty$ \cite{Negele}.

We then divide the fermionic field $\psi(r,t)$ into the two components 
$\psi_1(r,t)$ and $\psi_2(r,t)$ which reside on the forward and the 
backward parts of the time contour respectively. Since the interaction
part of the action is strictly local 
in time it may be rewritten as $S_{int}[\psi_1] -S_{int}[\psi_2]$ 
(the minus sign comes from the opposite direction of the time integral on 
the backward part of the contour) 
\begin{equation} 
S_{int}[\psi_i] \!=\!
\frac{-1}{2} \!\!\! \int \!\!\! dt\, d{\bf r}\, d{\bf r}' 
\overline\psi_i({\bf r})\overline\psi_i({\bf r}')
V_0({\bf r}-{\bf r}')\psi_i({\bf r}')\psi_i({\bf r})\,  ,
                                                              \label{e6}
\end{equation}
here $i=1,2$ and $V_0({\bf r}-{\bf r}')$ is a bare interaction potential.
We now introduce  two independent auxiliary bosonic fields $\hat
\phi_{1(2)}({\bf r},t)$ to decouple the two interaction terms by
the Hubbard--Stratonovich transformation. As a result one obtains for
the partition function    
\begin{mathletters}
                                                           \label{e7,8}
\begin{eqnarray} 
&&Z=\tilde {\cal N} \int {\cal D} \hat\Phi 
e^{ \frac{i}{2}\Tr\{\hat\Phi^T V_0^{-1}\sigma_3 \hat\Phi\}} 
\int {\cal D} \overline\Psi \Psi e^{iS[\overline\Psi,\Psi,\hat\Phi]}\, ,
                                                              \label{e7}
\\
&&S[\overline\Psi,\Psi,\hat\Phi]\!= \! 
\Tr\left\{ 
\overline\Psi [\hat G_0^{-1} - U_{dis}\sigma_3 
+\hat \phi_{\alpha}\hat\gamma^{\alpha}] \Psi 
\right \} .
                                                              \label{e8}
\end{eqnarray}
\end{mathletters}
Here we have introduced the following vector notations for the
fermionic doublet $\Psi$,  the bosonic doublet $\hat\Phi$, 
and two vertex matrices $\hat\gamma^\alpha$
\begin{mathletters}
\label{e9}
\begin{eqnarray} 
&&\Psi=\left(
\begin{array}{l}
\psi_1 \\
\psi_2
\end{array} \right)\, ;\hskip .7cm
\hat\Phi=\left(
\begin{array}{l}
\hat\phi_1 \\
\hat\phi_2
\end{array} \right)\, ; 
                                                             \label{e9a}\\
&&\hat\gamma^1=\left( \begin{array}{ll}
1&0 \\ 0&0 \end{array} \right)\, ;\,\,\,\,  
\hat\gamma^2=
\left( \begin{array}{rr}
0&0 \\ 0&-1 \end{array} \right)\, .
                                                              \label{e9b}
\end{eqnarray}
\end{mathletters}
The inverse matrix Green function stands for 
\begin{equation} 
\hat G_0^{-1} =
\left( \begin{array}{cc}
i\frac{\partial}{\partial t} + \frac{\nabla_r^2}{2m}&0 \\ 
0&- i\frac{\partial}{\partial t} - \frac{\nabla_r^2}{2m} 
\end{array} \right)\, .
                                                              \label{e10}
\end{equation}
The trace operation in Eq.~(\ref{e8}) and henceforth is understood to
be performed over the $2\times 2$ structure as well as over the time
and space variables.

\subsection{Disorder averaging}
\label{subsec.disorder}

The great advantage of the Keldysh technique is that the
normalization constant, $\tilde {\cal N}$ in  Eq.~(\ref{e7}) does not
depend on the realization of the disorder potential. 
Thus, the disorder averaging can be performed directly, without the need to
resort to the replica trick. Hereafter we employ the simplest 
model of the Gaussian, $\delta$--correlated disorder
\begin{equation}
\langle\ldots\rangle =
\int {\cal D} U_{dis} \dots
\exp\left \{-\pi\nu\tau\int d{\bf r} U^2_{dis}({\bf r}) \right\} ,
                                                              \label{d1}
\end{equation} 
where the  disorder strength is characterized by the elastic mean 
free time,  $\tau$; $\nu$ is the bare density of states at the Fermi
energy. Next, we perform the Gaussian integration over $U_{dis}$ in
Eq.~(\ref{e7})  and decouple the arising (non-local in time) quartic
interaction  by means of the Hubbard-Stratonovich transformation. 
Doing so, we obtain
\widetext
\top{-2.8cm}
\begin{mathletters}
\label{d2,3}
\begin{eqnarray} 
&&\left\langle e^{- i\, \Tr\{ \overline\Psi U_{dis}\sigma_3 \Psi\} } 
\right\rangle =
\exp \left\{ -(4\pi\nu\tau)^{-1}\!\! \int \!\! d{\bf r}
\left[ \int dt \overline\Psi({\bf r},t)\sigma_3 \Psi({\bf
r},t)\right]^2\right\} \\  
                                                               \label{d2}
&&= \int {\cal D} \hat Q 
\exp\left\{ -\int d{\bf r} dt dt'\left[ 
\frac{\pi\nu}{4\tau} \Tr\, \hat Q_{tt'}({\bf r})\hat Q_{t't}({\bf r})
+\frac{1}{2\tau} \overline\Psi({\bf r},t) \hat Q_{tt'}({\bf
r})\sigma_3 \Psi({\bf r},t') 
\right] \right\} .
                                                              \label{d3}
\end{eqnarray}
\end{mathletters}
\bottom{-2.7cm}
\narrowtext
\noindent
Here we have introduced the Hubbard-Stratonovich field $\hat{Q}$ which is a
matrix with indices both in  the Keldysh $2\times 2$ space and in the
time space. To ensure the convergence of the 
integral in Eq.~(\ref{d3}) the $\hat{Q}$-matrix is chosen to be Hermitian. 
After these transformations the fermionic functional integral in 
Eq.~(\ref{e7}) can be formally performed, leading to  
\begin{equation} 
\mbox{det} 
\left|
\hat G_0^{-1}+ \frac{i}{2\tau}\hat Q \sigma_3 
+\hat \phi_{\alpha}\hat\gamma^{\alpha}
\right| .
                                                              \label{d4}
\end{equation} 
As a result, the disorder averaged partition function takes the form 
\begin{mathletters}
\label{d5,6}
\begin{eqnarray} 
&&\langle Z \rangle =
 \int {\cal D} \hat\Phi\,  
e^{\,  \frac{i}{2}\Tr\{ \hat\Phi^T V_0^{-1}\sigma_3 \hat\Phi\} }
\int {\cal D} \hat Q\,  e^{iS[\hat Q,\hat\Phi]}\, ,
                                                              \label{d5} \\
&&iS[\hat Q,\hat\Phi]=\!
-\frac{\pi\nu}{4\tau}\,  \Tr\, \hat Q^2 \! + \!  
\Tr\ln \Big[
\hat G_0^{-1} \! +\!  \frac{i\hat Q \sigma_3 }{2\tau}  
\! +\! \hat \phi_{\alpha}\hat\gamma^{\alpha}
\Big]. 
                                                              \label{d6}
\end{eqnarray} 
\end{mathletters}
As before the trace operation is understood to be performed over 
the Keldysh and the time indices as well as over the coordinate space; 
the unessential normalization constant is omitted.

\subsection{Keldysh rotation}
\label{subsec.rotation}

In the notations introduced in Eqs.~(\ref{e9}) and (\ref{e10}) the
electron Green functions $\hat G $ are matrices in the $2 \times 2$
Keldysh space. Their components are not  independent and
satisfy certain general identities \cite{Keldysh,Rammer}. This
interdependence becomes most transparent if one introduces the rotated
Green functions $G$ denoted by the absence of the hat and defined as
\begin{equation}   
G \equiv L\sigma_3 \hat G  L^{\dagger}\, ,
                                                              \label{r3}
\end{equation}
where the unitary matrix $L$ is given by
\begin{equation} 
L=\frac{1}{\sqrt{2}}(\sigma_0 - i \sigma_2)=
\frac{1}{\sqrt{2}}
\left( \begin{array}{rr}
1&-1 \\ 1&1 \end{array} \right)\, .
                                                              \label{r2}
\end{equation}
As follows from the definition of the Keldysh Green function 
\cite{Keldysh,Rammer}, 
the rotated Green function has the following structure   
\begin{equation}   
G(t,t') = \left( \begin{array}{cc}
G^R(t,t') &G^K(t,t') \\ 0&G^A(t,t') \end{array} \right),
                                                              \label{r4}
\end{equation}
where $G^{R(A)}(t,t')$ vanish for $t\leq t'(t\geq t')$.
To pass to the rotated representation we introduce a new
Hubbard-Stratonovich field  
$Q$ which is related to the old one, $\hat{Q}$, by the following
unitary transformation 
\begin{equation}   
Q\equiv L\hat Q L^{\dagger}\, .
                                                              \label{r1}
\end{equation} 
We also introduce the rotated bare inverse Green function, $G_0^{-1}$,
expressed through $\hat G_0^{-1}$ of Eq.~(\ref{e10}) in a manner
consistent with Eq.~(\ref{r3})
\begin{equation} 
G_0^{-1}\equiv L \hat G_0^{-1}\sigma_3  L^{\dagger}
 =\left(
i\frac{\partial}{\partial t} + \frac{\nabla_r^2}{2m}  \right)\sigma_0 \, .
                                                              \label{r3a}
\end{equation}

It is also convenient to perform a linear transformation of the bosonic
doublet,  $\hat \Phi$, by introducing the symmetric and the antisymmetric
combinations  of the fields residing on the upper and the lower
branches of the contour ${\cal C}$   
\begin{equation}   
\begin{array}{c}
\phi_1=\frac{1}{2}(\hat \phi_1+\hat \phi_2) \, ;\\ 
\phi_2=\frac{1}{2}(\hat \phi_1-\hat \phi_2) \, .
\end{array} 
                                                              \label{r5}
\end{equation} 
Then the rotated vertex matrices for these new fields are 
$\gamma^{1(2)}=L(\hat\gamma^{1} \pm \hat\gamma^{2})\sigma_3
L^{\dagger}$  with the following explicit form 
\begin{equation} 
\gamma^1=\sigma_0= 
\left( \begin{array}{ll}
1&0 \\ 0&1 \end{array} \right)\, ;\,\,\,\,\, 
\gamma^2=\sigma_1=
\left( \begin{array}{ll}
0&1 \\ 1&0 \end{array} \right)\, .
                                                              \label{r8}
\end{equation}

Any classical external field takes on identical values on the two
branches 
of the contour and, hence, in the rotated basis has only the 
first symmetric component. The non--zero antisymmetric 
component may appear only as a virtual fluctuating field.
Below we shall sometimes refer to the first and second 
components of the bosonic fields as {\em classical} and 
{\em quantum} ones correspondingly. 
Since the presence of an external classical field does not change 
the basic fact that $Z=1$, any auxiliary source field should 
have a non--vanishing quantum component to generate an observable. 
We shall return to this observation in section \ref{subsec.sources}.

Utilizing the cyclic invariance of the trace operation we obtain the
following expression for the partition function, Eq.~(\ref{d5,6}), 
through the new variables $\Phi$ and $Q$ 
\begin{mathletters}
\label{r6,7}
\begin{eqnarray} 
&&\langle Z \rangle =
\int {\cal D} \Phi\,  
e^{ i\,\Tr\{ \Phi^T V_0^{-1}\sigma_1 \Phi\} }
\int {\cal D} Q\,  e^{iS[Q,\Phi]}\, ,
                                                              \label{r6}\\
&&iS[Q,\Phi] = \!
-\frac{\pi\nu}{4\tau}\,  \Tr\,  Q^2\!  + \! 
\Tr\ln \Big[
G_0^{-1} \! +\!  \frac{i}{2\tau} Q  
\! + \! \phi_{\alpha} \gamma^{\alpha}
\Big] .
                                                              \label{r7}
\end{eqnarray}
\end{mathletters}

\section{Non--Linear $\sigma$--Model}
\label{sec.smodel}

\subsection{Saddle point equation}
\label{subsec:sp}

We shall look now for a saddle point of the functional integral over
the $Q$-matrix in Eq.~(\ref{r6}). The aim is to find a stationary 
solution for a given realization of the slowly varying in space and
time fluctuating bosonic fields, $\Phi$. Calculating a variation of
the action,  Eq.~(\ref{r7}), over the $Q$-matrix, one obtains the
following equation for the saddle--point  matrix, $\underline Q =
\underline Q[\Phi]$,  
\begin{equation} 
\underline Q_{t,t'}(r)= \frac{i}{\pi\nu}
\left. \Big[
G_0^{-1}+ \frac{i}{2\tau} \underline Q  
+\phi_{\alpha} \gamma^{\alpha}
\Big]^{-1}  \right|_{r,r;t,t'} 
                                                              \label{p1}
\end{equation}    
We are unable to solve this equation exactly, therefore our goal will
be to  find its approximate solution, which is as close as possible to
the true  stationary point of the functional integral in
Eq.~(\ref{r6}). 
To execute this program, we first consider the case where 
$\Phi=0$. It is easy to check that in this case
\begin{equation} 
\Lambda_{t-t'}\equiv \underline Q[\Phi=0] = \frac{i}{\pi\nu}
\sum\limits_{p} G(p,t-t'),
                                                              \label{p1a}
\end{equation}
where the impurity averaged Keldysh Green function is
(cf. Eq.~(\ref{r4}))
\begin{eqnarray}   
&&G(p,\epsilon) = \left( \begin{array}{cc}
G^R(p,\epsilon) &G^K(p,\epsilon) \\ 0&G^A(p,\epsilon) \end{array} \right)
                                                             \label{p2}\\
&&=\left( \begin{array}{rr}
1&F_{\epsilon} \\ 0&-1 \end{array} \right)
\left( \begin{array}{cc}
G^R(p,\epsilon) & 0\\ 0&G^A(p,\epsilon) \end{array} \right)
\left( \begin{array}{rr}
1&F_{\epsilon} \\ 0&-1 \end{array} \right) ,
                                                              \nonumber
\end{eqnarray}
with 
\begin{mathletters}
\label{p3,4}
\begin{eqnarray}   
&&G^{R(A)}(p,\epsilon) = 
\left( \epsilon-\epsilon_p\pm i/(2\tau) \right)^{-1} \\ 
                                                           \label{p3}
&&G^K(p,\epsilon) = G^R F - F G^A \, .
                                                           \label{p4}
\end{eqnarray}
\end{mathletters}
The function $F$ defined by Eq.~(\ref{p4}) can be expressed through  the
single particle distribution function, $n(\epsilon)$,  as
$F(\epsilon)=1-2n(\epsilon)$. In equilibrium at temperature 
$T$ it is given by 
\begin{equation} 
F^{eq}_{\epsilon} = \tanh \frac{\epsilon}{2T}\, 
                                                           \label{p5}
\end{equation}
Substituting Eqs.~(\ref{p2}), (\ref{p3,4}) into Eq.~(\ref{p1a}) 
and performing the momentum summation, one obtains for the non--interacting 
($\Phi=0$) saddle point 
\begin{equation} 
\Lambda_{\epsilon} = 
\left( \begin{array}{cc}
\openone_{\epsilon}^R&2F_{\epsilon} \\ 0&-\openone_{\epsilon}^A \end{array} 
\right)\, ; \,\, 
\Lambda_{t-t'} = 
\left( \begin{array}{cc}
\delta_{t-t'-0} & 2F_{t-t'} \\ 0& -\delta_{t-t'+0} \end{array} \right)\, .
                                                              \label{p6}
\end{equation}
We have introduced here the retarded and the advanced unities,
$\openone^{R(A)}_{\epsilon}$, which should be understood as Fourier
transforms of infinitesimally shifted $\delta$--functions. 
This particular form of the Green function is a result of
the approximation that the single-particle DOS  
is independent of the energy, $\epsilon$. 
In reality it does depend on $\epsilon$, and the retarded (advanced)
components of $\Lambda (\epsilon)$ are analytic functions of energy in
the upper (lower) half plane which do depend on energy on 
the scale  of order of the Fermi energy, $\epsilon_F$. 
Therefore the infinitesimally shifted $\delta$-functions in
Eq.~(\ref{p6}) should be understood as $\delta_{t \mp 0} = 
f_\pm (t) \Theta(\pm t)$, where $\Theta( t)$ is the Heavyside
function, and $f_\pm (t)$ are functions that are highly peaked for
$|t| \lesssim \epsilon_F^{-1}$ and satisfy the  normalization
condition $\int_0^{\pm \infty} dt f_\pm (t)=1$.  This
high--energy regularization is important to remember in calculations to
avoid spurious unphysical constants. In particular, for 
obvious reasons  
\begin{equation} 
 \begin{array}{ll}
\openone^R_{t-t'} {\cal M}^R_{t',t}=0\, ; \\ 
\openone^A_{t-t'} {\cal M}^A_{t',t}=0\, ,
\end{array} \,  
                                                              \label{p7}
\end{equation} 
where ${\cal M}^{R(A)}_{t',t}$ is an arbitrary retarded (advanced)
matrix in  the time space.

Substituting Eq.~(\ref{p6}) into Eq.~(\ref{p1}) with $\Phi=0$, it is
easy to see that $ \underline Q = \Lambda $ solves the
non--interacting saddle--point equation for {\em any} function
$F_{\epsilon}$. (The simplest way to check it is 
to use the decomposition  Eq.~(\ref{p2}).) This is natural, since any
distribution function is allowed for the non--interacting electron
gas. We shall see below how the interaction effects drive the system
towards the equilibrium distribution, Eq.~(\ref{p5}). 

Let us now include a finite $\phi_{\alpha}(r,t)$ into Eq.~(\ref{p1}). 
To this end we notice that this equation can be still solved exactly
for the  particular case of spacially uniform  realizations of the
boson field, $\phi_{\alpha}=\phi_{\alpha}(t)$. This is obvious since
such a field may be gauged out resulting in 
\begin{equation} 
\underline Q_{t,t'}[\Phi(t)]  =
e^{i \int\limits^{t}dt \phi_{\alpha}(t)\gamma^{\alpha} }
\Lambda_{t-t'} 
e^{-i \int\limits^{t'}dt \phi_{\alpha}(t)\gamma^{\alpha} }
                                                              \label{p8}
\end{equation}  
The validity of this solution can be verified by acting with the
operator  $[G_0^{-1}+ i/(2\tau) \underline Q  +\phi_{\alpha}(t)
\gamma^{\alpha}]$ on both sides of Eq.~(\ref{p1}) and utilizing the
fact that $\Lambda_{t-t'}$ solves Eq.~(\ref{p1}) with $\Phi=0$.
We also rely on the commutativity of the vertex matrices,
$[\gamma^1,\gamma^2]=0$, in writing the solution in the form
of Eq.~(\ref{p8}).

We now consider the case where the bosonic fields $\phi_\alpha$ are
slowly (compared to the mean free path $l$) varying in space. In
analogy with Eq.~(\ref{p8}) we shall look for an approximate solution
of  Eq.~(\ref{p1}) in the form of a local (in time and space) 
gauge transformation of $\Lambda$ 
\begin{equation} 
\underline Q_{t,t'}(r) =
e^{i k_{\alpha}(r,t)\gamma^{\alpha}} 
\Lambda_{t-t'} 
e^{-i k_{\alpha}(r,t')\gamma^{\alpha}}\, , 
                                                              \label{p9}
\end{equation}  
where $k_{\alpha}=k_{\alpha}[\Phi]$  
is a certain {\em linear} functional of the fields 
$\Phi$, whose specific form is to be determined to satisfy 
Eq.~(\ref{p1}) in the best possible way. 

To proceed we introduce a new Hubbard-Stratonovich field $\tilde Q$
which is related to the old one, $Q$, by the gauge transformation 
\begin{equation} 
Q_{t,t'}(r) =
e^{i k_{\alpha}(r,t)\gamma^{\alpha}} 
 \tilde Q_{t,t'}(r)  
e^{-i k_{\alpha}(r,t')\gamma^{\alpha}}\, . 
                                                              \label{p9a}
\end{equation}   
Substituting this definition  into the action,
Eq.~(\ref{r8}), and using the invariance of the trace 
under a cyclic permutation of operators, we can rewrite the action as
\begin{eqnarray} 
&&iS[\tilde Q,\Phi]=
-\frac{\pi\nu}{4\tau}\,  \Tr \tilde Q^2 +  
                                                           \label{p12}\\
&&\Tr\ln  \Big[
G_0^{-1} + C-\sum_\alpha \frac{(\nabla k_\alpha)^2}{2m} + 
\frac{i}{2\tau} \tilde Q 
\Big] \, ,           
                                                            \nonumber
\end{eqnarray}    
where we have introduced the notation
\begin{equation} 
C(r,t) \equiv   
(\phi_{\alpha}-\partial_t k_{\alpha}- {\bf v}_F\nabla k_{\alpha}) 
\gamma^{\alpha}\, ,                   
                                                            \label{p13}
\end{equation} 
with the Fermi velocity, ${\bf v}_F=-i\nabla_r/m$. 
To find the approximate saddle point of the form Eq.~(\ref{p9}) we
substitute $\tilde Q = \Lambda + \delta \tilde Q$ into Eq.~(\ref{p12})
and require terms linear in $\delta \tilde Q$ to vanish. 
In doing so we neglect the diamagnetic term, $(\nabla
k_{\alpha})^2/2m$, since it is quadratic in $k_{\alpha} $ (and hence
in $\Phi$) and is also smaller than $C$ in the  parameter $q/p_F \ll 1$, 
where $q$ is the characteristic momentum scale of variation of $\Phi$. 
%
%
As a result we obtain the following equation
\begin{equation} 
-\pi\nu  \Lambda_{t',t}+ 
i \left[G-G C G + GCGCG -\ldots \right]_{t',t}(r,r) = 0 \, .
                                                            \label{p14}
\end{equation} 
The first two terms in this expression cancel, according 
to Eq.~(\ref{p1a}). The freedom of choosing the $K[\Phi]$ functional is
not sufficient to  
cancel all the terms in this expansion. We thus concentrate 
on the term which is linear in $\Phi$ and $K$: 
\begin{eqnarray} 
&&\sum\limits_p G(p_+,\epsilon_+) C(q,\omega) G(p_-,\epsilon_-) 
= \pi\nu\tau \times
                                                            \label{p15}\\
&& \left[  
(\phi_{\alpha} \!+\! i\omega k_{\alpha})
(\gamma^{\alpha} \!-\! \Lambda_{+}\gamma^{\alpha}\Lambda_{-})
\!-\! Dq^2k_{\alpha} 
(\Lambda_{+}\gamma^{\alpha} \!-\! \gamma^{\alpha}\Lambda_{-})
\right], 
                                                            \nonumber
\end{eqnarray}  
where $p_{\pm}= p\pm q/2,\, \epsilon_{\pm}= \epsilon\pm\omega/2$ and 
$\Lambda_{\pm} = \Lambda_{\epsilon_{\pm} }$. To derive  
Eq.~(\ref{p15}) one may employ the following useful representation of 
the Keldysh Green function 
\begin{eqnarray} 
&&G(p,\epsilon)\equiv [G_0^{-1}+\frac{i}{2\tau}\Lambda_{\epsilon}]^{-1}
                                                           \label{p16} \\
&& =
\frac{1}{2} G^R(p,\epsilon)(\sigma_0 +\Lambda_{\epsilon}) +   
\frac{1}{2} G^A(p,\epsilon)(\sigma_0 -\Lambda_{\epsilon})\, .
                                                           \nonumber 
\end{eqnarray} 
Only $\sum_p G^R G^A$ and $\sum_p G^R {\bf v}_F G^A$ contribute to Eq.~(\ref{p15}). 
Multiplying Eq.~(\ref{p15}) by $\Lambda_{\epsilon_+}$ from the 
left one obtains the following matrix condition for the vanishing of
the linear  term in Eq.~(\ref{p14}) 
\begin{equation} 
Dq^2k_{\alpha}
(\Lambda_{+}\gamma^{\alpha}\Lambda_{-} \!-\!  \gamma^{\alpha})\! +\!
(\phi_{\alpha} \! + \! i\omega k_{\alpha})
(\Lambda_{+}\gamma^{\alpha}\! - \! \gamma^{\alpha}\Lambda_{-}) 
\! =\! 0.
                                                            \label{p17}
\end{equation}
To cancel $(1,1)\, , (2,2)$ and $(2,1)$ components of the matrix on the 
l.h.s. of this equation the functional $K$ should satisfy 
\begin{equation} 
(Dq^2+i\omega) k_{2}(q,\omega) + \phi_{2}(q,\omega)=0\, .
                                                            \label{p18}
\end{equation} 
Provided this equality is obeyed, the condition to cancel the
Keldysh $(2,1)$ component  on the l.h.s. of Eq.~(\ref{p17}) is 
\begin{equation} 
(Dq^2-i\omega) k_{1} - \phi_{1} =
-2Dq^2 k_{2} \, 
\frac{1- F_{\epsilon_+}F_{\epsilon_-}}{F_{\epsilon_+}-F_{\epsilon_-}}.
                                                            \label{p19}
\end{equation} 
This equation can not be in general satisfied, since its r.h.s.
contains explicit dependence on $\epsilon$, whereas the l.h.s. is supposed
to be $\epsilon$-independent. This happens, because the trial saddle
point solution, Eq.~(\ref{p9}), is too restrictive.  In particular, we
demanded that it may be obtained from $\Lambda$ by the rotation which is
local in time. As a result, the  functional $K$ depends only on
differential energy $\omega$ and not on ``center of mass'' energy
$\epsilon$. This restriction is in an apparent contradiction with
Eq.~(\ref{p19}). There is, however, an important particular case when
Eq.~(\ref{p19}) may be solved. This is the case of thermal
equilibrium, where the fermionic distribution function is given by
Eq.~(\ref{p5}). As a result   
\begin{equation} 
\frac{1- F^{eq}_{\epsilon_+}F^{eq}_{\epsilon_-}}
{F^{eq}_{\epsilon_+}-F^{eq}_{\epsilon_-}}=
\coth\frac{\omega}{2T}\equiv B^{eq}_\omega\, ,
                                                            \label{p20}
\end{equation}
where $B^{eq}_\omega$ stands for the equilibrium bosonic distribution
function.  
Thus, the matrix equation (\ref{p17}) for the functional $K[\Phi]$
can be resolved in equilibrium. The result may be written in a short  
form as 
\begin{equation}   
{\cal D}^{-1}(q,\omega) K(q,\omega) =  
\Pi_{\omega}^{-1}\Phi(q,\omega)\, .
                                                              \label{p21}
\end{equation}
We have introduced here the following  bosonic matrix propagators 
in the $2\times 2$ Keldysh space 
\begin{equation}   
{\cal D}(q,\omega) = 
\left( \begin{array}{cc}
{\cal D}^K(q,\omega) & {\cal D}^R(q,\omega) \\
{\cal D}^A(q,\omega) &         0
\end{array} \right)\, , 
                                                              \label{p22}
\end{equation}
with
\begin{mathletters}
\label{p23}
\begin{eqnarray} 
&&{\cal D}^{R(A)}(q,\omega)=  (Dq^2 \mp i\omega)^{-1} , \\ 
&&{\cal D}^K(q,\omega) =  
B_\omega({\cal D}^R(q,\omega) - {\cal D}^A(q,\omega) )\, ;         
\end{eqnarray} 
\end{mathletters}
and 
\begin{equation}   
\Pi_{\omega}= -i \omega {\cal D}(q=0,\omega) = 
\left( \begin{array}{cc}
2 B_\omega  & 1^R_\omega  \\ 
-1^A_\omega &   0  
\end{array} \right)\, . 
                                                              \label{p22a}
\end{equation}
The superscript ``{\em eq}''  denoting equilibrium in the bosonic
distribution has been omitted for brevity.

Eqs.~(\ref{p9}) and (\ref{p21}) complete the task of finding 
the approximate saddle point, $\underline Q = \underline Q[\Phi]$, for
any given realization of fields $\Phi$. On this solution  we are
able to cancel only the term linear in $\Phi$ in the expansion 
Eq.~(\ref{p14}). This guarantees only that the terms like 
$\Phi\delta \tilde Q$ will not appear in the expansion of the action
around the saddle point given by Eqs.~(\ref{p9}), (\ref{p21}). Terms
like $\Phi^2 \delta \tilde Q$  may (and
will) arise in such  expansion. We shall see later, that it is 
precisely these terms that are responsible for the divergent
Altshuler--Aronov corrections to conductivity\cite{Altshuler85}. 
The ability to avoid
$\Phi\delta \tilde Q$ terms is, strictly speaking, limited only to the
thermal  equilibrium. For an out of equilibrium situation such terms
reappear and  require some care (see Section \ref{sec.kinetic}).

The influence of the external potential $\Phi$ on
the electron dynamics (and hence on the Green function) is
two-fold \cite{LandauQM}: i) It changes the particle trajectory and  
ii) It changes the phase of the electron wave function. 
The first effect is proportional to the electric field ${\bf E}=-\nabla
\Phi$ and is small for the long wave length spatial configurations of
$\Phi$. The second effect, however, requires no actual electric fields. 
It is proportional to $\Phi$ itself, rather than $\nabla
\Phi$, and is akin to the Aharonov-Bohm effect. It changes the phase
but not the amplitude of the wave function and can be taken into
account in the Eikonal approximation. The second effect exceeds the 
first one for the long wave length fluctuations of the potential,
therefore, it is especially important in the presence of the 
long range Coulomb
interactions. The approximation to the saddle
point, Eq.~(\ref{p9}), is similar to the Eikonal
approximation. It is designed to account for the phase effect of the 
slow fluctuations of the potential $\Phi$. Note that  
the phase $K$ enters the saddle point equation only through
its total time derivative along the trajectory of a particle,  
$d/dt =\partial_t+ {\bf v}_F\nabla$ (cf Eq.~(\ref{p13})). 
If we demand that $C$ vanish we obtain the standard
Eikonal equation \cite{LandauQM} for the action $K$ of the particle
moving with a given velocity ${\bf v}$ in an external field
$\Phi$. Unfortunately, the ansatz Eq.~ (\ref{p9}) is too restrictive to
nullify $C$ for particles of every velocity ${\bf v}$.  
Eventually all the particles in the Fermi sea
interfere to produce the Green function $\underline Q_{t,t'}(r)$. 
Equation (\ref{p9}) approximately accounts for the phase interference
between particles moving along different trajectories.  Since the
particle dynamics is diffusive this leads to the diffusive relation 
(\ref{p21})--(\ref{p23}) between the external potential $\Phi$ and the
phase $K$. 
As will be clear below the choice of the saddle point in the form 
Eq.~(\ref{p9}) considerably  simplifies the subsequent calculations.
In particular, it eliminates completely the family of super-divergent
diagrams which cancel in the traditional treatment \cite{Finkelstein83} 
after sometimes tedious calculations.

\subsection{Effective action}
\label{subsec:effac}

To formulate an effective low energy theory in terms of the
fluctuating fields $\tilde Q$ and $\Phi$ we need to examine the
fluctuations around the saddle point.  The fluctuations
of $\tilde Q$ fall into two general
classes\cite{Wegner79,Efetov82,Finkelstein83}: i) massive, with the mass
$\propto 1/\tau$  and ii) massless, those on which the action depends only
very weakly. The fluctuations along the massive modes can be
integrated out 
in the Gaussian approximation and lead to insignificant
renormalization of various parameters in the action. The massless, or
Goldstone, modes describe diffusive motion of the electrons. The
fluctuations of the $\tilde Q$-matrix along these massless modes are not
small and should be treated carefully.  The 
Goldstone modes can be parameterized by the 
$\tilde Q$-matrices satisfying
a certain nonlinear constraint \cite{Wegner79,Efetov82,Finkelstein83}. 

To identify the relevant Goldstone modes consider the first term 
in Eq.~(\ref{p12}). The saddle point is given by Eqs.~(\ref{p9}) and
(\ref{p21}) satisfies 
\begin{equation} 
\tilde Q^2 = \left( \begin{array}{cc} 
\openone_\epsilon^R & 0\\
0 & \openone_\epsilon^A
\end{array}
 \right)
\, ,                        
                                                            \label{q5}
\end{equation}  
and the first term in Eq.~(\ref{p12}) vanishes. The fluctuations 
of $\tilde Q$ which do not satisfy Eq.~(\ref{q5}) are massive. 
The massless modes are generated by rotations of the saddle point 
and can be parameterized as \cite{Wegner79,Efetov82,Finkelstein83}
\begin{equation} 
\tilde Q  = {\cal T}^{-1}\Lambda {\cal T}\, .  
                                                            \label{q7}
\end{equation} 
The parameterization of the rotation matrices ${\cal T}$ must ensure
the convergence of the functional integration over the matrices 
given by Eq.~(\ref{q7}). Below we only assume that such a
parameterization exists, whereas the concrete form of ${\cal T}$ is 
not important for what follows. 

One way of parameterizing the rotations is to write 
${\cal T}=\exp \{ W/2 \}$,
where, without loss of generality,  $W \Lambda =-\Lambda W$. 
Expanding Eq.~(\ref{p12}) to the second order in $W$ and neglecting
for a moment the term arising due to e--e interactions 
it is easy to establish that in the diffusive regime the relevant
fluctuations must satisfy the condition
\begin{equation} 
W_{\epsilon,\epsilon'}\neq 0\, ,\,\,\,  
\mbox{only if} \,\,\, |\epsilon| , |\epsilon'| < 1/\tau\, . 
                                                              \label{q1}
\end{equation}
Namely, all effective degrees of freedom are concentrated in the
narrow energy  strip of the width  $1/\tau\ll \epsilon_F$ near the Fermi
energy.  Therefore the matrices ${\cal T}$ differ from unity 
only in the narrow region of energies defined by Eq.~(\ref{q1}).
For  this reason the gauge transformation 
${\cal U}_{t,t'}(r)=\exp\{-ik_\alpha(t,r)\gamma_\alpha\}\delta(t-t')$ 
in Eq.~(\ref{p9a}) 
can not be incorporated into a redefinition of ${\cal T}$ 
and should be carried out explicitly. Indeed, being diagonal in time
indices, the matrix  ${\cal U}_{t,t'} $ spreads over the entire energy
space and, thus, can not be reduced to a disturbance, which is close
to the Fermi shell. Physically, this describes the fact the 
low--wavenumber scalar potential $\Phi(q,t)$ shifts the entire
electronic band and not only the energy strip given by Eq.~(\ref{q1}). 
It is essential to follow the
variations  of the electron spectrum all the way down to the bottom of the
band to respect the charge neutrality imposed by  the Coulomb
interactions.  As we shall see below, once the phase factors 
in Eq.~(\ref{q7}) have been taken into account, the residual 
interactions may be regarded as being short-range 
without loss of generality.  
 
Substituting Eq.~(\ref{q7}) into Eq.~(\ref{p12}),   and
retaining only the universal ($\tau$ -- independent) terms in  the
expansion of the logarithm, we obtain for the $\tilde Q$ action  
\begin{eqnarray} 
&&iS[\tilde Q,\Phi]= 
i\nu \Tr \{(\Phi-i\omega K)^T \sigma_1 (\Phi+i\omega K) \} -
                                                             \label{q8}\\
&&\frac{\pi\nu}{4} \left[ 
D  \Tr \{ {\bf \partial}_r\tilde Q\}^2 +
4i \Tr \{(\epsilon +(\phi_\alpha+i\omega 
k_{\alpha})\gamma^{\alpha}) \tilde Q \}
\right] \, ,
                                                                \nonumber  
\end{eqnarray} 
where we have introduced the long derivative 
\begin{equation} 
{\bf \partial}_r\tilde Q \equiv 
\nabla \tilde Q +i [\nabla k_{\alpha}\gamma^{\alpha}, \tilde Q]\, .       
                                                            \label{q9}
\end{equation}  
A few comments are in order regarding Eq.~(\ref{q8}). 
First, it is restricted to $\tilde Q$ which satisfy Eq.~(\ref{q5}).
The last two terms, containing $\tilde Q$, conventionally originate from
$\sum_p {\bf v}_F G^R {\bf v}_F G^A$ and $\sum_p G^{R(A)}$ 
combinations in the
expansion of the logarithm. On the other hand, 
the first term in the r.h.s. of
Eq.~(\ref{q8}) originates from $\sum_p G^R G^R$ and $\sum_p G^A G^A$
combinations.  These terms should be retained since, as was mentioned
above, the matrix $\phi_\alpha(\epsilon-\epsilon')\gamma^{\alpha}$ is
not restricted to the $1/\tau$ shell near the Fermi energy. 
To derive this term we employed the fact  
that for any physical fermionic distribution function 
\begin{equation} 
F_{\epsilon\rightarrow \pm \infty} \rightarrow \pm 1\, .        
                                                            \label{q9a}
\end{equation}  
Finally, the terms like $\sum_p {\bf v}_F G^R {\bf v}_F G^R$, 
although non--vanishing, cancel 
against the diamagnetic term. 

Employing the explicit form of the long derivative, 
Eq.~(\ref{q9}), and the relation 
between the $K$ and $\Phi$ fields, Eq.~(\ref{p21}), 
one finally obtains for the 
the partition function
\top{-2.8cm} 
\widetext
\begin{equation} 
\langle Z \rangle =
\int\!\!  {\cal D} \Phi\,  
\exp\{ i\,\Tr\{ \Phi^T V^{-1} \Phi\} \}
\int \!\!{\cal D} \tilde Q\,  
\exp \Big\{ 
iS_0[\tilde Q]+iS_1[\tilde Q,\nabla K]+iS_2[\tilde Q,\nabla K]
\Big\}\, , 
                                                              \label{q10}
\end{equation}
where $S_l, \,  l=0,1,2$ contain  $\nabla K$ in the $l-th$ power and 
are given by  
\begin{mathletters}
\label{q11,12,13}
\begin{eqnarray} 
&&iS_0[\tilde Q]=
-\frac{\pi\nu}{4} 
\left[ D  \Tr \{ \nabla \tilde Q\}^2 +  4i\,  \Tr \{\epsilon \tilde  Q
\} \right]\, ; 
                                                              \label{q11} \\
&&iS_1[\tilde Q,\nabla K]=
-i\pi\nu  
\left[ D  \Tr \{ \nabla k_\alpha\gamma^{\alpha} \tilde Q\nabla \tilde Q \} +  
\Tr \{(\phi_\alpha+i\omega  k_\alpha)\gamma^{\alpha} \tilde Q\} \right]\, ;
                                                              \label{q12} \\
&&iS_2[\tilde Q,\nabla K]=
\frac{\pi\nu D}{2}  
\left[
\Tr \{ \nabla k_\alpha\gamma^{\alpha} \tilde Q 
       \nabla k_\beta \gamma^{\beta } \tilde  Q \} -  
\Tr \{ \nabla k_\alpha\gamma^{\alpha} \Lambda 
       \nabla k_\beta \gamma^{\beta } \Lambda \} 
\right]\, .
                                                              \label{q13} 
\end{eqnarray} 
\end{mathletters}
\bottom{-2.7cm}
\narrowtext
\noindent
The effective interaction matrix $V$ is nothing but 
the screened interaction in the RPA approximation 
\begin{equation}   
V(q,\omega) = 
\left( V_0^{-1}(q)\sigma_1 +P_0(q,\omega) \right)^{-1}\,  ,
                                                              \label{q13a}
\end{equation}
where $P_0(q,\omega)$ is the bare density--density correlator. 
It has a typical form of a bosonic correlator  in the Keldysh 
space  
\begin{equation}   
P_0(q,\omega) = 
\left( \begin{array}{cc}
0               & P_0^A(q,\omega) \\
P_0^R(q,\omega) & P_0^K(q,\omega)   
\end{array} \right)\,  ,
                                                              \label{q14}
\end{equation}
with 
\begin{mathletters}
\label{q15}
\begin{eqnarray} 
&&P_0^{R(A)}(q,\omega)=  
\nu \frac{Dq^2}{D q^2 \mp i\omega}  \, ,   
                                                         \label{q15a}    \\
&&P_0^K(q,\omega) =  
B_\omega(P_0^R(q,\omega) - P_0^A(q,\omega) )\, .         
                                                              \label{q15b}
\end{eqnarray}  
\end{mathletters}
To derive Eqs.~(\ref{q10})--(\ref{q15}) we had to add and subtract the term 
$\Tr \{ \nabla k_\alpha\gamma^{\alpha} \Lambda 
\nabla k_\beta\gamma^{\beta} \Lambda\}$ and employed the equation  
\begin{equation}   
\int\limits_{-\infty}^{+\infty} \!\!\! d\epsilon \, 
\Tr \{ \gamma^{\alpha}\gamma^{\beta} - 
\gamma^{\alpha} \Lambda_{\epsilon_+} 
\gamma^{\beta}  \Lambda_{\epsilon_-}   \}=
4\omega \left( \Pi_\omega^{-1} \right)^{\alpha\beta}\, .
                                                              \label{q16}
\end{equation}
Here $\epsilon_{\pm}= \epsilon\pm \omega/2$ and 
matrices $\Lambda$ and $\Pi$ are
defined by the Eqs.~(\ref{p6}) and (\ref{p22a}) 
correspondingly. Eq.~(\ref{q16})
based on the following relations between bosonic 
and fermionic distribution functions
\begin{eqnarray} 
&&\int\limits_{-\infty}^{+\infty} \!\!\! d\epsilon\, 
(F_{\epsilon_+} - F_{\epsilon_-}) = 2\omega\, ;
                                                              \label{q17}\\
&&\int\limits_{-\infty}^{+\infty} \!\!\! d\epsilon\, 
(1- F_{\epsilon_+}F_{\epsilon_-}) = 2\omega B_\omega \, .
                                                              \label{q18}
\end{eqnarray}   
The last equation is obviously satisfied in the thermal equilibrium. For a 
non--equilibrium situation it should be considered as a 
definition of $B_\omega$.

Eqs.~(\ref{q10})--(\ref{q15}) together with Eq.~(\ref{p21}) 
constitute an effective
non--linear $\sigma$--model for interacting disordered electron gas. 
The model consists of two interacting fields: matrix field $\tilde Q$,
obeying the  
non--linear constraint Eq.~(\ref{q5}), and the bosonic vector 
field $\Phi$ (or equivalently $K$). As will be apparent later, the 
$\tilde Q$--field describes  fluctuations of the 
quasi--particle distribution function, whereas $\Phi$ 
(or $K$) represents 
propagation of electromagnetic fields through the media.   
The following 
sections are devoted to the analysis of this model and 
calculation of various 
physical quantities on the basis of the model.

\section{Applications of the Formalism}
\label{sec.applic}


\subsection{Single particle Green function}
\label{subsecGF}

In this section we shall show how the developed formalism can be used 
for the calculation of the average single--particle Green 
function at coinciding spatial points. This quantity is defined as 
\begin{equation} 
\hat {\cal G}_{i,j}(t-t') = 
-i \langle\langle 
\psi_i(r,t) \overline \psi_j(r,t') 
\rangle\rangle\, ,
                                                          \label{z1}
\end{equation}
where $\langle \langle \ldots  \rangle \rangle $ denotes both the 
quantum and the disorder averaging. It is convenient to 
apply the Keldysh rotation, Eq.~(\ref{r3}), and define 
\begin{equation} 
{\cal G}(t-t') = L\sigma_3 \hat {\cal G}(t-t') L^{\dagger}  \, .
                                                          \label{z2}
\end{equation}
Such  Green function arises e.g. in calculations of the tunneling DOS, or 
shot noise power. To evaluate it one may introduce  a source term 
in Eq.~(\ref{e3}), directly coupled to a bilinear combination of the fermion 
operators. Following the same algebra as above one finds that the source 
field enters into the logarithm in Eq.~(\ref{r7}). Differentiating finally 
with respect to the source and putting it to zero one obtains for the
Green function  
\begin{eqnarray} 
{\cal G}(t-t') =
&&\int\!\!  {\cal D} \Phi\,  
e^{ i\,\Tr\{ \Phi^T V_0^{-1}\sigma_1 \Phi\} } \times 
                                                              \label{z3}\\
&&\int \!\!{\cal D} Q\,  
e^{ i S[Q,\Phi] } 
\left. \Big[ G_0^{-1}+ \frac{i}{2\tau}  Q  
+\phi_{\alpha} \gamma^{\alpha} \Big]^{-1} \right|_{{\bf r},{\bf r};t,t'} \, .
                                                              \nonumber
\end{eqnarray}
We shall evaluate the integral over the $Q$ matrix by the saddle 
point approximation, neglecting both the massive and the massless 
fluctuations around the stationary point.
Then, according to Eq.~(\ref{p1}), the pre-exponential factor is simply 
$-i\pi\nu\underline Q_{t,t'}$. At the saddle point $\underline Q$ is given by 
Eq.~(\ref{p9}). Transforming the action, $S[Q,\Phi]$, in the way 
it was done in section \ref{subsec:effac}, one obtains for the Green 
function in the saddle point approximation
\begin{equation} 
{\cal G}\! =\!  -i\pi\nu \!\!
\int\!\!\!  {\cal D} \Phi\,  
e^{ i\,\Tr\{ \Phi^T V^{-1} \Phi\} }
e^{i k_{\alpha}(t)\gamma^{\alpha}} 
\Lambda_{t-t'} 
e^{-i k_{\alpha}(t')\gamma^{\alpha}}\,  .
                                                              \label{z4}
\end{equation} 
Since $K$ is the linear functional of $\Phi$, given by Eq.~(\ref{p21}), the 
remaining functional integral is Gaussian. Employing Eqs.~(\ref{p21}) and 
(\ref{q14})--(\ref{q15}), one obtains for the correlator of the $K$ fields 
(averaged over the fluctuations of $\Phi$)
\begin{mathletters}
\label{z5}
\begin{eqnarray}   
&&\left\langle k_{\alpha}(q,\omega) k_{\beta}(-q,-\omega) \right\rangle_{\Phi}
=  {i\over 2} {\cal V}_{\alpha\beta}(q,\omega) \, ;
                                                              \label{z5a}\\
&&{\cal V}(q,\omega)\! = \! 
{\cal D}(q,\omega)\Pi^{-1}_{\omega} V(q,\omega)\! 
\left(\Pi^{-1}_{-\omega}\right)^T \!\! {\cal D}^T(-q,-\omega)\, .
                                                               \label{z5b}
\end{eqnarray}
\end{mathletters}
The Keldysh matrix ${\cal V}$ has the familiar structure of a bosonic
propagator 
\begin{equation}   
{\cal V}(q,\omega) = 
\left( \begin{array}{cc}
{\cal V}^K(q,\omega) & {\cal V}^R(q,\omega) \\
{\cal V}^A(q,\omega) &         0
\end{array} \right)\,  ,
                                                              \label{z7}
\end{equation}
with 
\begin{mathletters}
\label{z8}
\begin{eqnarray} 
&&{\cal V}^{R(A)}(q,\omega) \!=\!  
\frac{-1}{(Dq^2 \mp i\omega)^2}
\left( \frac{1}{V_0}\! + \! 
\frac{\nu Dq^2}{D q^2 \mp i\omega} \right)^{-1}  ,  
                                                             \label{z8a}    \\
&&{\cal V}^K(q,\omega) =  
B_\omega({\cal V}^R(q,\omega) - {\cal V}^A(q,\omega) )\, .         
                                                              \label{z8b}
\end{eqnarray}  
\end{mathletters}
One may recognize that this propagator precisely corresponds to the screened 
Coulomb interaction line dressed by two diffusons at the vertices. 
Thus, the role of the $K$--field is to take into account 
automatically both the RPA-screened interactions and its vertex
renormalization by diffusons. 

To calculate the functional integral, Eq.~(\ref{z4}), we write the phase 
factors as 
\begin{eqnarray} 
e^{\pm i k_{\alpha}\gamma^{\alpha}}  = 
&&{1\over2}\left( e^{\pm i(k_1+k_2)}+ e^{\pm i(k_1-k_2)}\right) \gamma^1 +
                                                              \label{z10}\\
&&{1\over2}\left( e^{\pm i(k_1+k_2)}- e^{\pm i(k_1-k_2)}\right) \gamma^2\, 
                                                              \nonumber
\end{eqnarray} 
and perform the Gaussian integration according to Eq.~(\ref{z5}).
The result may be conveniently expressed in the following form
\begin{equation} 
{\cal G}(t) = -i\pi\nu 
\sum\limits_{\alpha\beta=1}^{2}
\left( \gamma^{\alpha}\Lambda_{t}\gamma^{\beta} \right)
{\cal B}_{\alpha\beta}(t) \,  .
                                                              \label{z11}
\end{equation} 
where the fictitious propagator  ${\cal B}$ has the standard bosonic structure
(as e.g.  Eq.~(\ref{z7})) with
\widetext
\top{-2.8cm}
\begin{mathletters}
\label{z12}
\begin{eqnarray} 
{\cal B}^{R(A)}(t) &&= 
\frac{1}{2} e^{{i\over2}({\cal V}^K(t)-{\cal V}^K(0))}
\left(e^{{i\over2}{\cal V}^{R(A)}(t)} - 
e^{-{i\over2}{\cal V}^{R(A)}(t)}\right)  \, ,  
                                                             \label{z12a}  \\
{\cal B}^K(t) &&=  
\frac{1}{2} e^{{i\over2}({\cal V}^K(t)-{\cal V}^K(0))} 
\left(e^{{i\over2}({\cal V}^{R}(t)-{\cal V}^{A}(t)) } +          
e^{-{i\over2}({\cal V}^{R}(t)-{\cal V}^{A}(t)) }  \right)\, .
                                                             \label{z12b}
\end{eqnarray}  
\end{mathletters}
\bottom{-2.7cm}
\narrowtext
\noindent 
The $\langle K K^T \rangle$ propagator, ${\cal V}$, defined by 
Eqs.~(\ref{z7}),(\ref{z8}) is taken at coinciding spatial 
points 
\begin{equation}   
{\cal V}(t)= \int\!   \frac{d\omega}{2\pi} e^{-i\omega t}
\sum\limits_{q}{\cal V}(q,\omega)\, .
                                                              \label{z14}
\end{equation}

The electron Green function must satisfy several important
requirements:   the tunneling  DOS, $\nu(\epsilon)$, 
which is defined as 
\begin{equation}   
\nu(\epsilon) = {i\over 2\pi }
\left( {\cal G}^R(\epsilon)- {\cal G}^A(\epsilon)\right)\, 
                                                              \label{z14c}
\end{equation}
must be a positive definite quantity. In addition,  
in thermal equilibrium the  $R$, $A$ and $K$ components of the 
bosonic and fermionic propagators are related by the
fluctuation-dissipation theorem (FDT). Below we demonstrate that our
approximation, Eqs.~(\ref{z11}) and (\ref{z12}), for the Green function 
satisfies these requirements. For this purpose it is convenient 
to rewrite identically Eq.~(\ref{z11}) in the following form 
\begin{equation} 
  {\cal G}^{>(<)}(t) = -i\pi\nu \Lambda^{>(<)}_{t}
{\cal B}^{>(<)}(t) \,  ,                                \label{z14a}
\end{equation}
where 
\begin{mathletters}
\label{z16,17}
\begin{eqnarray} 
{\cal B}^{R}(t)-{\cal B}^{A}(t) &&=
 {\cal B}^{>}(t) -{\cal B}^{<} (t)   \, ,  
                                                             \label{z16a}  \\
{\cal B}^K(t) &&=  
 {\cal B}^{>}(t) +{\cal B}^{<} (t)  \, .
                                                               \label{z17a}
\end{eqnarray} 
\end{mathletters}
The $>$ and $<$ components of the fermionic Green functions are related to 
the $R$, $A$ and $K$ in the same manner.  
>From Eqs.~(\ref{z16,17}) and Eqs.~(\ref{z12}) we obtain
\widetext
\top{-2.8cm}
\begin{equation}   
{\cal B}^{>(<)}(\omega) = {1\over 2} e^{-{i\over2}{\cal V}^K(0)} 
\int\!\! dt\,  e^{i\omega t}
\exp\left\{\! {i\over2} \! 
\int\!\! \frac{d\omega'}{2\pi}\, e^{-i\omega' t}
\sum\limits_{q}({\cal V}^R(q,\omega')- {\cal V}^A(q,\omega'))
\left(\coth\frac{\omega'}{2T}\pm 1 \right) \right\}  \, .
                                                              \label{z18a}
\end{equation}
\bottom{-2.7cm}
\narrowtext
\noindent 
According to the FDT the equilibrium bosonic and fermionic Green functions
in the frequency representation satisfy the following relations 
\begin{mathletters}
\label{z19g}
\begin{eqnarray} 
 {\cal B}^{>}(\omega)&=&\exp\{\omega/T\}{\cal B}^{<} (\omega)   \, ,  
                                                          \label{z19a}  \\
 {\cal G}^{>}(\epsilon) &=& - \exp\{\epsilon/T\}{\cal G}^{<} (\epsilon)  \, .
                                                               \label{z19b}
\end{eqnarray} 
\end{mathletters}

It is not difficult to see that if any pair of bosonic Green functions
 $B^{>}(t)$ and $B^{<}(t)$ satisfies 
Eq.~(\ref{z19a}) then for any analytic function $f(z)$ the pair 
$f^>(t)\equiv f(B^{>}(t))$ and $ f^<(t)\equiv f(B^{<}(t))$ 
also satisfies it. 
Indeed,  
\begin{equation}
f^{>(<)}(\omega) = \int dt e^{i\omega t} f\left( 
\int\frac{d\omega'}{2\pi} B^{>(<)}(\omega')e^{-i\omega' t} \right)
\, .         
                                                              \label{z20n}
\end{equation}  
Expanding $f$ on the r.h.s. of this equation in the Taylor series  
and performing the $t$ integration, we see that in each order of
the expansion $f^{>}(\omega)=\exp(\omega/T)f^{<}(\omega)$. 
One can also  check that if ${\cal G}^{>(<)}$ and 
${\cal B}^{>(<)}$ satisfy the FDT, Eq.~(\ref{z19g}), then so do 
the functions  $\tilde{\cal G}^{>(<)}$ defined as
\begin{equation}
\tilde{\cal G}^{>(<)}(t)= {\cal G}^{>(<)}(t){\cal B}^{>(<)}(t)
\, .         
                                                              \label{z21n}
\end{equation}  
Noting that the arguments in the exponential in  Eq.~(\ref{z18a})
obviously satisfy the FDT, Eq.~(\ref{z19a}), we conclude that 
${\cal B}^{>(<)}(\omega)$, Eq.~(\ref{z18a}),  
and the approximate Green function, Eq.~(\ref{z11}),  
satisfies it as well.

To establish the positive definiteness of the tunneling density of
states, Eq.~(\ref{z14c}), we first show that 
${\cal B}^{>(<)}(\omega)$ in Eq.~(\ref{z18a}) is positive definite. 
Indeed, $\exp[-i{\cal V}^K(0)]$ is real as can be seen from Eqs.~(\ref{z8}).
It is also not difficult to see that each Fourier component 
of the argument of the exponential in (\ref{z18a}) is positive definite.
All the coefficients in the Taylor series of the exponential are positive
and, since the Fourier transform of a product is given by the
convolution of {\em positively defined} Fourier transforms, we 
conclude that the l.h.s. of (\ref{z18a}) is positive definite. 
We next use Eqs.~(\ref{z16a}) and (\ref{z19a}) to 
rewrite the tunneling density of states as 
\begin{equation}
\nu(\epsilon) = {i\over 2\pi }
 {\cal G}^>(\epsilon) (1+e^{-\epsilon/T}) 
\, ,         
                                                              \label{z22n}
\end{equation}  
where ${\cal G}^>(\epsilon)$ is given by Eq.~(\ref{z14a}). 
Since $\Lambda ^>_\epsilon \geq 0$ we immediately see from
Eq.~(\ref{z14a}) that the tunneling density of states is positive.

\begin{figure}
\vglue 0cm
\hspace{0.01\hsize}
\epsfxsize=0.9\hsize
\epsffile{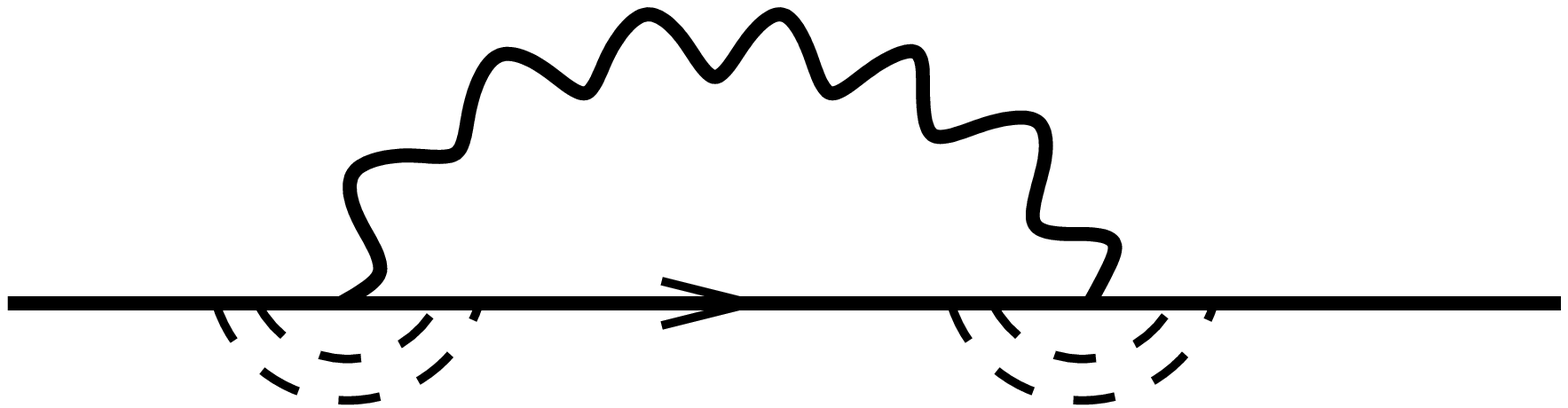}
\refstepcounter{figure} \label{gf:fig}
{\small FIG.\ \ref{gf:fig}
Lowest order interaction correction to the single particle Green
function. The wavy line here denotes the RPA-screened Coulomb 
interaction. The impurity-dressed single particle Green functions are
depicted by solid  lines and the double dashed lines
represent diffusons. \par}
\end{figure}

In equilibrium, it is convenient 
to write the DOS through the Keldysh Green function employing the FDT 
\begin{equation}   
\nu(\epsilon) = \frac{\nu}{\tanh \epsilon/(2T)}   
\int\! dt\,  e^{i\epsilon t}\, F_{t} \,
{\cal B}^K(t)\, ,
                                                              \label{z23}
\end{equation}
As was proven above, Eqs.~(\ref{z22n}) and (\ref{z23}) are equivalent. 
One can then express 
${\cal B}^K(t)$ through ${\cal B}^{>(<)}(t)$, where the latter are 
conveniently rewritten as 
\widetext
\top{-2.8cm}
\begin{equation}   
{\cal B}^{>(<)}(t) = {1\over 2} 
\exp \left\{\! 
\int\!\! \frac{d\omega}{2\pi}\, 
\left(\coth \frac{\omega}{2T} (1 - \cos\omega t) \pm  i\sin\omega t \right)
\Im \sum\limits_{q}{\cal V}^R(q,\omega)
\right\}
                                                              \label{z24}
\end{equation}
\bottom{-2.7cm}
\narrowtext
\noindent  
Expanding this expression to the first order in the interaction, ${\cal V}$, 
and substituting into Eq.~(\ref{z23}) one recovers  the 
Altshuler--Aronov result for the zero--bias anomaly \cite{Altshuler85}.
This perturbative result corresponds to the diagram drawn in Fig.\ref{gf:fig}.

We shall restrict ourselves to the analysis of the non--perturbative result,
Eq.~(\ref{z23}), (\ref{z24}), only at $T=0$. 
Noting that for $T=0$,   $F_t=(i\pi t)^{-1}$, one obtains 
\begin{eqnarray}   
\nu(\epsilon) \!&=&\! \frac{\nu}{\pi} \!
\int\!\! dt\, \frac{\sin|\epsilon| t}{t} \,
\exp\! \left\{\! 
\int\limits_0^{\infty}\!\! \frac{d\omega}{\pi}\, 
\Im \sum\limits_{q}{\cal V}^R(\omega)(1 \!-\! \cos\omega t)\!
\right\} 
                                                        \nonumber \\
&\times& \cos\left\{\!
\int\limits_0^{\infty}\!\! \frac{d\omega}{\pi}\, 
\Im \sum\limits_{q}{\cal V}^R(\omega)\sin\omega t\right\} \, .
                                                          \label{z25}
\end{eqnarray}
In the two dimensional case Eq.~(\ref{z8a}) with $V_0(q)=2\pi e^2/q$
leads to 
\begin{eqnarray}   
&&\int\limits_0^{\infty}\!\! \frac{d\omega}{\pi}\, 
\Im \sum\limits_{q}{\cal V}^R(q,\omega)
\left( \begin{array}{c}
1 - \cos\omega t   \\
\sin\omega t
\end{array} \right) 
                                                            \label{z26}\\
&& = 
-\frac{1}{8\pi^2 g} \left\{  \begin{array}{c}
\ln t/\tau \ln t\tau\omega_0^2 +2\gamma \ln t\omega_0 \\
\pi \ln t\omega_0
\end{array} \right.    \, , 
                                                              \nonumber
\end{eqnarray}
where $g=\nu D$ is the conductance;
$\omega_0=D\kappa^2$ and $\kappa=2\pi e^2 \nu$ is the inverse screening 
radius; $\gamma=.577\ldots$ is the Euler constant. 
Since we have neglected the fluctuations of $\tilde Q$, 
we have missed corrections 
of order $g^{-1} \ln t/ \tau$ (in $d=2$), therefore Eq.~(\ref{z25}) can  
only be trusted for $\epsilon$ not too small, such that 
$(8\pi^2 g)^{-1} \ln(\epsilon\tau)^{-1} \ll 1$. 
However $\ln^2 t/\tau$ terms have been accounted for correctly 
by our procedure. If in addition $g^{-1} \ln \omega_0\tau\ll 1$, 
the time integral in Eq.~(\ref{z25})
may be performed by the stationary point method, resulting in 
\begin{equation}   
\nu(\epsilon) =\nu 
\exp \left\{\! 
-\frac{1}{8\pi^2 g}
\ln (|\epsilon|\tau)^{-1} \ln \tau\omega_0^2/|\epsilon| 
\right\}\, .
                                                              \label{z27}
\end{equation}
Theoretically, however,  $g^{-1} \ln \omega_0\tau$, need not be small.
In that case the stationary point integration should be somewhat modified and 
terms $\propto \ln t\omega_0 $ should be retained. 

We have achieved a  non--perturbative resummation of anomalously divergent, 
$\propto \ln^2 \epsilon\tau$, 
terms for a single particle Green function. 
The non--perturbative expression for the DOS, 
essentially arises from the gauge non--invariance of the single particle 
Green function. The calculations above are in essence the
``Debaye--Waller'' factor\cite{Finkelstein94} 
due to almost pure gauge fluctuations of electric fields, cf. Eq.~(\ref{z4}).
Gauge invariant characteristics (such as e.g. conductivity) do not carry phase
factors and therefore are not affected by the interactions on this level of 
accuracy (fluctuations of $\tilde Q$ should be  retained). In perturbation 
theory this fact is reflected by the cancellation of diagrams without 
diffusons (apart from those which renormalize vertices) \cite{Altshuler85}. 
In our formulation such terms never appear since the phase factors cancel 
along any closed loop diagram.   

The gauge physics of anomalous corrections to the 
DOS was first realized by
Finkel'stein \cite{Finkelstein83,Finkelstein94}, who obtained a 
non--perturbative result 
similar to ours. Nazarov \cite{Nazarov89} and later 
Levitov and Shytov \cite{Levitov} obtained the same result 
(in imaginary time) by semiclassical reasoning. Kopietz \cite{Kopietz98} 
recently reinstated it, stressing the role of phase fluctuations. 
The analogous expression for the zero--dimensional case has also been 
known for  
some time \cite{Devoret91,Glazman91,Kamenev96}. We believe that we provide 
its first consistent derivation using the $\sigma$--model. Unlike the previous 
approaches, the Keldysh technique provides the answer directly in real 
time and finite temperature. This enables us to circumvent the 
tedious analytical continuation procedure.

\subsection{Shot noise}
\label{subsecSN}

In this subsection we shall use the results obtained in 
subsection \ref{subsecGF} to calculate the power spectrum of current 
noise  through a tunneling contact between a clean metal and 
a dirty metal film. 
The power spectrum of current noise is given by
\begin{equation}
S(\omega) =  \int dt e^{i\omega t} \langle \langle 
\hat{I}(t)\hat{I}(0)+ \hat{I}(0)\hat{I}(t) 
\rangle \rangle\, .         
                                                              \label{sn1}
\end{equation}  
Here the current operator in the tunneling approximation is
given by $\hat{I}(t)=iT a^\dagger(t,r_0)b(t,r_0) - 
iT^*b^\dagger(t,r_0)a(t,r_0)$, and $b(t,r_0)$ and $a(t,r_0)$ are
electron annihilation operators at the position of the contact 
in the dirty film and in the clean metal respectively. 
Below all the fermion operators and the Green functions are 
taken at the point of the tunneling contact $r_0$, and 
we omit the position argument for brevity. 
Using the expression for the current operator we can rewrite
Eq.~(\ref{sn1})  as 
\widetext
\top{-2.8cm} 
\begin{equation}
S(\omega) =  |T|^2 \!\!\int \!\! dt e^{i\omega t}  
\Big[  
{\cal G}_a^<(-t){\cal G}_b^>(t) +{\cal G}_a^>(t) {\cal G}_b^<(-t) + 
{\cal G}_a^<(t){\cal G}_b^>(-t) + {\cal G}_a^>(-t){\cal G}_b^<(t)
\Big] \, ,         
                                                              \label{sn2}
\end{equation}
where ${\cal G}_a$  and ${\cal G}_b$ are Green functions for the 
clean metal and for the dirty film respectively. 
We assume that the voltage $V$ is applied across the contact.
To the lowest order in the tunneling matrix element, the Green
functions under these conditions are equilibrium, except that 
the chemical potentials in the two metals differ by $eV$. Therefore 
in the lowest order in the tunneling amplitude we can express the 
power spectrum of current noise through the equilibrium Green
functions. Expressing them through DOS with the aid of the FDT and
utilizing the fact that for the clean metal DOS, $\nu_a$, is 
independent of energy $\epsilon$ one obtains
\begin{equation}
S(\omega) =   2\pi \nu_a |T|^2\!\! \int \!\! 
d\epsilon \, \nu_b (\epsilon)  
\Big[  n(\epsilon) [2-n(\epsilon +\omega -eV)-
n(\epsilon -\omega -eV)] 
+  [1-n(\epsilon)]\left( n(\epsilon +\omega -eV)+ 
n(\epsilon -\omega -eV) \right) \Big]
\, ,        
                                                      \label{sn3} 
\end{equation}
\bottom{-2.7cm}
\narrowtext
\noindent 
where $n(\epsilon)=[1+\exp(\epsilon/T)]^{-1}$ is the Fermi function.

Setting $V=0$ in Eq.~(\ref{sn3}) we obtain the power spectrum of the
equilibrium current noise in the contact $S_0(\omega)$. 
The excess noise is given by the difference $\delta S(\omega)= 
 S(\omega)- S_0(\omega)$. The noise power is a symmetric function 
of frequency, and at zero temperature reduces to 
\begin{equation}
\delta S(\omega>0) = 2\pi \nu_a |T|^2 
\left[ 
 \int\limits_{-|\omega-eV|}^{\omega+eV}\!\!\! 
d \epsilon\, \nu_b(\epsilon)
- \int\limits_{-\omega}^{\omega}d \epsilon \, \nu_b(\epsilon) 
\right] \, ,         
                                                              \label{sn4}
\end{equation}

At zero frequency the shot noise is proportional to the total
current. This is natural, since in the lowest order in the
tunneling amplitude the electrons pass through the contact 
extremely rarely and, therefore, can be viewed as noninteracting. The
role of interactions reduces to modification of the density of the
available states. The cusp present at zero temperature in the noise
power spectrum for noninteracting electrons at $\omega=0$ and
$\omega=eV$ is washed out because DOS vanishes at $\epsilon=0$.

\subsection{External fields and auxiliary sources}
\label{subsec.sources}

In some sense our previous manipulations leading to 
Eqs.~(\ref{q10}), (\ref{q11,12,13}) were no more than 
a complicated representation of unity. This is so since, according to
the basic idea of the Keldysh technique,  
the partition function, $Z$, is identically equal to unity. To make
the entire  
construction meaningful one should introduce auxiliary source
fields, which enable one to compute various observables. We shall do
it in parallel with introducing external classical fields. Since we
shall mostly discuss the conductivity, we'll use  the vector
potential  ${\bf A}(r,t)$ as an example \cite{foot2}. 
Other fields (e.g. the scalar
potential) may be introduced in a similar way.   
We introduce a doublet in the rotated Keldysh frame
\begin{equation} 
{\bf A}(r,t)=
\left( \begin{array}{c}
{\bf a}_1(r,t)\\
{\bf a}_2(r,t)
\end{array} \right)\, ,
                                                              \label{u1}
\end{equation}
which is related by the usual transformation (cf. Eq.~(\ref{r5})) with the 
two fields $\hat a_{\alpha}(r,t)$ residing on the two branches of the
time contour. The vector potentials enters the fermionic Hamiltonian
through the  long spatial derivatives, 
$\nabla_r \rightarrow \nabla_r +i\hat{\bf a}_{\alpha} \hat \gamma^{\alpha}$.
The classical external vector potential is the same on the two
branches of the contour and hence it is described by the symmetric
component, ${\bf a}_1(r,t)$, only, whereas ${\bf a}_2=0$. 
In this case the generating  function still equals to unity 
\begin{equation} 
Z[{\bf a}_1, {\bf a}_2=0] =1\, .
                                                              \label{u2}
\end{equation}  
To obtain a non--trivial generating function one has to introduce a
quantum  component of the source field, ${\bf a}_2(r,t)$. This component
does not have a  classical meaning and thus has to be nullified at the
end of the  calculations. Its presence however is essential for
generating observables. One can easily check that the
current density defined as 
\begin{equation} 
{\bf j}={e\over 2m i}  
\left\langle 
{1\over 2}\sum\limits_{i=1,2}\left\{ 
\overline \psi_i (\nabla + i  {\bf a}_1) \psi_i - 
(\nabla - i  {\bf a}_1)\overline \psi_i \psi_i\right\}
\right\rangle_{\psi} 
                                                              \label{u2a}
\end{equation}
is given by \cite{foot4}
\begin{equation} 
{\bf j}(r,t)= -{e\over 2i} \left. 
\frac{\delta Z[{\bf A}]}{\delta {\bf a}_2(r,t)} 
\right|_{{\bf a}_2=0}
\, .
                                                              \label{u3}
\end{equation}  
We restrict ourselves to the case of longitudinal vector potentials
only. Taking into account the fact that the external electric field is  
given by $-i\omega {\bf a}_1(q,\omega)/e$, one obtains that the linear
response conductivity is given by the retarded component of the  
current--current correlator 
\begin{equation} 
\sigma(q,\omega) = {e^2\over i \omega} \Sigma^{2,1}(q,\omega)\, ,
                                                              \label{u4}
\end{equation}
where 
\begin{equation} 
\Sigma^{\alpha,\beta}(q,\omega) = {1\over 2i}\, \left. 
\frac{\delta^2 Z[{\bf A}]}{\delta a_\beta({\bf q},\omega) \delta
a_\alpha(-{\bf q},-\omega)}  
\, \right|_{{\bf A}=0}
\, .
                                                              \label{u4a}
\end{equation}
Here we have omitted the vector indices of $a$ using its longitudinal
character. 
In general, any response function is given by the (2,1) component of
the appropriate  bosonic correlator. The structure of the theory
guarantees that this is a retarded function
(cf. e.g. Eq.~(\ref{q14})).

In the presence of an external vector potential, ${\bf A}$, the trial
saddle point,  Eq.~(\ref{p9}), is shifted. Noting that ${\bf A}$
enters the action always in the  
combination $\nabla K + {\bf A}$, one finds that the condition for the  
optimal $K$ is given by  Eq.~(\ref{p17}), with the substitution 
$Dq^2K\rightarrow Dq^2 K+ i D(qA)$ \cite{foot3}. 
Solving this equation in a manner
it was done in section \ref{subsec:sp}, one obtains that
Eq.~(\ref{p21}) should  
be modified as 
\begin{equation}   
{\cal D}^{-1}(q,\omega) K(q,\omega) =  
\Pi_{\omega}^{-1}\Phi(q,\omega)
-i D\sigma_1 (qA(q,\omega))\, ,
                                                              \label{u5}
\end{equation}
where bosonic propagators ${\cal D}(q,\omega)$ and 
$\Pi_\omega=-i\omega{\cal D}(q=0,\omega)$ are defined by 
Eqs.~(\ref{p22}) and (\ref{p22a}). In solving  Eq.~(\ref{p17}) with the
external field we still assumed that the fermionic distribution
function is the equilibrium one. This is a legitimate procedure in 
linear response. The generalization to the
non--equilibrium case is discussed in section \ref{sec.kinetic}. 
After disregarding the massive modes and expanding the logarithm, 
one obtains  Eq.~(\ref{q8}), 
with the long derivative modified as 
\begin{equation} 
\partial_r\tilde Q \equiv 
\nabla_r \tilde Q +
i [(\nabla k_{\alpha}+{\bf a}_{\alpha})\gamma^{\alpha}, \tilde Q]\, 
                                                            \label{u6}
\end{equation}
and $K$  given by Eq.~(\ref{u5}). 
Since $\gamma^1=1$, any static external field, ${\bf a}_1(r)$, appears
to be decoupled from $\tilde Q$. This reflects the fact that diffusons are
not coupled to a static magnetic field. On the other hand, even space
and time independent quantum component, ${\bf a}_2$, is coupled to $\tilde Q$. 
A little algebra shows that 
\begin{mathletters}
\label{u7,8}
\begin{eqnarray}
\nabla K + {\bf A} &&= -i{\bf q} {\cal D}\Pi^{-1}
\left( \Phi + (q A)\omega/q^2 \right)\, ; 
                                                            \label{u7}\\
\Phi + i\omega K &&= Dq^2 {\cal D} \sigma_1
\left( \Phi + (q A)\omega/q^2 \right)\, .                      
                                                            \label{u8}
\end{eqnarray}
\end{mathletters}
With these expressions and the long derivative given by Eq.~(\ref{u6}), 
one can rearrange Eq.~(\ref{q8}) to obtain the average generating 
function in the following form 
\widetext
\top{-2.8cm}
\begin{equation} 
\langle Z[A] \rangle =
\int\! {\cal D}\Phi 
\exp \Big\{ 
i\Tr \{ \Phi^T V_0^{-1}\sigma_1 \Phi  +
[\Phi + (q A)\omega/q^2]^T P_0 [\Phi + (q A)\omega/q^2]\} 
\Big\}
\int\! {\cal D} \tilde Q
\exp \left\{ \sum\limits_{l=0}^{2} i S_{l}[\tilde Q,\nabla K+{\bf A}] 
\right\} \, ,
                                                              \label{u9}
\end{equation}
\bottom{-2.7cm}
\narrowtext
\noindent 
where  the action $S_l, \,\,\, l=0,1,2$ is  given by 
Eqs.~(\ref{q11,12,13}) and the bare polarization operator, 
$P_0(q,\omega)$, is given by Eq.~(\ref{q15}). 
By virtue of Eqs.~(\ref{u7,8}) the entire action is expressible 
through the combination $\Phi + (q A)\omega/q^2$, which is  
proportional to the (gauge invariant) electric field 
$\nabla\Phi+\partial_t {\bf A}$. This fact immediately guarantees that the 
continuity equation is satisfied to all orders in the perturbation theory.
Indeed, one could introduce the external scalar potential, $\varphi$, 
which enters the action always as $\Phi+\varphi$ (apart from 
the bare interaction term, $V_0$). Then,
due to the fact that $Z=Z[\varphi + (q A)\omega/q^2]$,
the charge density,  
$\rho=(2i)^{-1} \delta Z/\delta\varphi_2$, and current density, 
${\bf j}=-(2i)^{-1} \delta Z/\delta {\bf a}_2$, has to be related by 
\begin{equation} 
(\nabla {\bf j})+\partial_t \rho=0\, .
                                                              \label{u10}
\end{equation}
The corresponding variation with respect to the  classical components,
$\varphi_1$ and  
${\bf a}_1$, guarantees continuity at each branch of the contour separately 
even in the presence of non--zero auxiliary quantum fields.  
As a result of continuity the exact relations between current--current and 
density--density correlators holds 
\begin{equation} 
P(q,\omega) = {q^2\over \omega^2}\, \Sigma(q,\omega)\, .
                                                              \label{u10a}
\end{equation}

At the saddle point, $\tilde Q=\Lambda$, one has $S_l[\Lambda,\nabla
K+{\bf A}]=0$. Thus, neglecting the fluctuations of $\tilde Q$, one
obtains  for the RPA  generating function 
\begin{eqnarray} 
&&\langle Z_{RPA}[A] \rangle =
\exp \left\{ iA^T\frac{\omega^2}{q^2}P_0 A \right\} \times 
                                                            \label{u11}\\
&&\int{\cal D}\Phi
\exp \left\{  
i\Tr\{ \Phi^T V^{-1} \Phi +2\Phi^T\frac{\omega}{q}P_0 A \}
\right\}  \, .
                                                            \nonumber
\end{eqnarray}
Performing finally the Gaussian integration, one finds 
\begin{equation} 
\langle Z_{RPA}[A] \rangle =
\exp\left\{ 
i\Tr\{ A^T \frac{\omega^2}{q^2} P_{RPA} A  \} 
\right\} \, , 
                                                              \label{u12}
\end{equation}
where $P_{RPA}(q,\omega)$ is the RPA  screened density--density
(polarization) correlator, which is given by 
\begin{equation} 
P_{RPA}(q,\omega)= P_0-P_0[V_0^{-1} + P_0]^{-1}P_0 = 
[P_0^{-1} +V_0]^{-1}\, 
                                                              \label{u14}
\end{equation}  
and has the structure of a bosonic correlator 
\begin{equation} 
P(q,\omega)=\left( \begin{array}{cc}
0             &  P^A(q,\omega)\\
P^R(q,\omega) &  P^K(q,\omega)
\end{array} \right)\, .
                                                              \label{u13}
\end{equation} 
The $\omega^2/q^2$ factor in Eq.~(\ref{u12}) reflects the relation
between the density--density and the longitudinal component of the
current--current  correlators, Eq.~(\ref{u10a}). 
The fact that the (1,1) component of $P$ (as well as of any other 
bosonic correlator) vanishes is a manifestation of the normalization 
condition, Eq.~(\ref{u2}). Employing Eq.~(\ref{u4}) one obtains for the 
conductivity in the  RPA 
\begin{equation} 
\sigma_{RPA}(q,\omega)= e^2\nu D 
\frac{-i\omega}{Dq^2(1+\nu V_0(q)) - i\omega}  \, ,
                                                              \label{u15}
\end{equation}

One is usually interested in the {\em irreducible} part of  the 
density--density (or current--current) correlators, 
which describes the linear response to the {\em total} or internal 
field and not to the external field as discussed above. 
The relation between the irreducible part, $P_{irr}$, and the total
$P$ is exactly the same as between the bare, $P_0$, and $P_{RPA}$,
Eq.~(\ref{u14}), 
\begin{equation} 
P_{irr}(q,\omega) = 
[P^{-1}(q,\omega) - V_0(q)]^{-1}\, .
                                                              \label{u16}
\end{equation}   
Shifting the integration variable 
$\Phi\rightarrow \Phi-(qA)\omega/q^2$ in Eq.~(\ref{u9}) and differentiating 
twice with respect to $A$ one obtains an exact  relation for the 
polarization operator 
\begin{equation} 
P(q,\omega) = V_0^{-1} +2i 
V_0^{-1}  \langle \Phi(q,\omega)\Phi^T(-q,-\omega) \rangle
V_0^{-1}   \, ,
                                                              \label{u17}
\end{equation}  
where $\langle \Phi\Phi^T\rangle$ is an exact propagator (averaged 
with respect to the full action, Eq.~(\ref{u9})). Employing
Eq.~(\ref{u16}) one finds 
\begin{equation} 
P_{irr} = {i \over 2} 
\left( \langle \Phi \Phi^T \rangle \right)^{-1} - V_0^{-1}\, .
                                                              \label{u18}
\end{equation}

If one is interested in the response to a uniform external field,
$q=0$, the expressions may be further simplified. Noting that for the
Coulomb interaction $V_0^{-1}(q=0)=0$ and employing Eq.~(\ref{u10a})
and relation between $\Phi$ and $K$, Eq.~(\ref{p21}), one obtains
\begin{equation} 
\Sigma_{irr}(q=0,\omega) = {i \over 2} 
\Big( \langle \nabla K(\omega) \nabla K^T(-\omega) 
\rangle 
\Big)^{-1} \, .
                                                              \label{u19}
\end{equation}
Unlike in the calculations of the single--particle 
Green function, only $\nabla K$ and never $K$ itself appears in 
calculations of gauge invariant quantities. This allows one to
consider  a universal  limit of strong interactions
$V_0^{-1}(q)\rightarrow  0$. In this limit  it is convenient to change
the integration variable from $\Phi$ to $\nabla K$ (although formally
it is a  vector it has only a longitudinal component and hence number
of variables is conserved). In the new variables the Gaussian weight
is given by $i\Tr\{\nabla K^T q^{-2}{\cal V}^{-1}\nabla K \}$, where
${\cal V}$ is defined by Eqs.~(\ref{z7})--(\ref{z8}). In the 
universal    limit one has 
\begin{equation} 
{\cal V}^{-1}(q,\omega) \rightarrow 
-\nu D q^2 \,  {\cal D}^{-1}(q,\omega)\, ,
                                                              \label{u20}
\end{equation}
where the diffusion propagator ${\cal D}$ is defined 
by Eqs.~(\ref{p22}), (\ref{p23}). Finally, one obtains for the 
action in terms of $\nabla K$ 
\begin{equation} 
\langle Z \rangle \! = \! 
\int\!\! {\cal D}\nabla K\, 
e^{ -i\nu D \Tr\{ \nabla K^T {\cal D}^{-1} \nabla K\} } 
\!\! \int\!\!\!  {\cal D} \tilde Q\, 
e^{ \sum\limits_{l=0}^{2} i S_{l}[\tilde Q,\nabla K] } \, .
                                                              \label{u21}
\end{equation}
Eqs.~(\ref{u19}) and (\ref{u21}) constitute a complete framework for 
calculations of gauge--invariant response functions. 
Neglecting $\tilde Q$--fluctuations one finds 
$\Sigma_{irr}(q=0,\omega)= \nu D i \omega \Pi^{-1}_\omega$, which leads to 
the Drude conductivity, $\sigma=e^2\nu D$.   
Fluctuations of $\tilde Q$ and $\nabla K$ lead to weak--localization and 
interaction corrections. Note that unlike in the case of the DOS 
(section \ref{subsecGF}), fluctuations of $\nabla K$ alone, with
$\tilde Q=\Lambda$,  do not lead to any corrections to linear response. 
This is a direct consequence of gauge invariance of
linear response functions. Only combined fluctuations of  $\nabla K$
and $\tilde Q$, discussed in  the next section,  renormalize the Drude
conductivity.

\section{Fluctuation Effects}
\label{sec.fluct}

\subsection{$\tilde Q$--matrix parameterization}
\label{subsec.param}

As was discussed in section \ref{subsec:effac} the massless fluctuations of 
the $\tilde Q$--matrix can be parameterized as 
\begin{equation} 
\tilde Q=\exp\{-W/2 \} \Lambda \exp\{W/2 \}\, ,                         
                                                            \label{l2}
\end{equation} 
where 
\begin{equation} 
W\Lambda+ \Lambda W = 0  \, .                         
                                                            \label{l3}
\end{equation} 
Employing Eq.~(\ref{p2}), one obtains that the general form of $W$, 
which satisfies the condition (\ref{l3}) is
\begin{eqnarray} 
W &=&
\left( \begin{array}{cc}
1 &  F \\
0 & -1
\end{array} \right)
\left( \begin{array}{cc}
          0 &  w \\
\overline w &  0
\end{array} \right)
\left( \begin{array}{cc}
1 &  F \\
0 & -1
\end{array} \right)
                                                          \label{l4}\\
&=&
\left( \begin{array}{cc}
F \overline w   &  -w +F\overline w F\\
-\overline w    &    -\overline w F
\end{array} \right)                        
                                                          \nonumber 
\end{eqnarray} 
where $\overline w$ and $w$ are arbitrary Hermitian matrices in the
time space. Below we shall thus understand the functional integration over  
$\tilde Q$ as integration over Hermitian  $\overline w$ and
$w$. Notice that $\tilde Q$ itself (as well as the Green function $G$)
appears to be non--Hermitian. It means that the ``contour'' of
integration in the $\tilde Q$ space is deformed from being pure
Hermitian to pass through the non--Hermitian saddle point.  
As it will be apparent later, the physical meaning of $w$ is a
deviation of the fermionic distribution function, $F$, from its
stationary value. At the same time, $\overline w$ has no classical
interpretation. To a large extent it plays the role of the quantum
counterpart of $w$, which appears  
only as the internal line in the diagrams. 

One may expand now the action,  Eqs.~(\ref{q11,12,13}), in
powers  
of $\overline w$ and $w$. The expansion of the non--interacting action, 
$iS_0[\tilde Q]$  starts from the second order, which has a familiar
diffusive  
structure 
\begin{equation} 
iS_0^{(2)}[W]=
\frac{\pi\nu}{2}
\overline w_{\epsilon_1\epsilon_2} 
\left[ -D  \nabla^2 +  i(\epsilon_1-\epsilon_2) \right]
w_{\epsilon_2\epsilon_1} \, .
                                                              \label{l5} 
\end{equation}
As a result the bare propagator of the $\tilde Q$--matrix fluctuations is given by
\begin{eqnarray} 
\left\langle 
w_{\epsilon_2\epsilon_1}(q)\overline w_{\epsilon_3\epsilon_4}(-q)
\right\rangle_{W} &=&
-\frac{2}{\pi\nu}\, 
\frac{\delta_{\epsilon_1\epsilon_3}\delta_{\epsilon_2\epsilon_4} }
{Dq^2+ i(\epsilon_1-\epsilon_2) } 
                                                            \label{l5a}\\
&=&
-\frac{2\delta_{\epsilon_1\epsilon_3}\delta_{\epsilon_2\epsilon_4}}{\pi\nu} 
D^{A}(q,\epsilon_1-\epsilon_2) \, .
                                                            \nonumber
\end{eqnarray}
The higher order terms describe non--linear interactions of diffusons 
with the vertices having the structure of Hikami boxes. One can easily work 
out this expansion in the Keldysh formalism. We shall not do it here, since
our main focus is on the interaction effects. Substituting 
$\delta \tilde Q^{(1)}=[\Lambda, W]/2$ 
into $iS_{1}[\tilde Q,\nabla K]$, one obtains in 
the first order in $W$
\widetext
\top{-2.8cm}
\begin{equation} 
iS_1^{(1)}[W,\nabla K]=
-{i\pi\nu \over 2}  
\Tr \left \{ \left[ 
D \nabla^2 k_\alpha (\Lambda \gamma^{\alpha}\Lambda - \gamma^{\alpha})  
+  
(\phi_\alpha+i\omega  k_\alpha) 
(\gamma^{\alpha}\Lambda- \Lambda \gamma^{\alpha}) 
\right] W \right\}\, .
                                                              \label{l6} 
\end{equation}
In equilibrium $iS_1^{(1)}[W,\nabla K]=0$. Indeed the r.h.s. of  
Eq.~(\ref{l6}) coincides with equation (\ref{p17}), which was 
used to determine  the $K$ functional. In equilibrium we were able 
to solve  Eq.~(\ref{p17}) by an appropriate choice of $K$. 
This was precisely the motivation behind 
looking for the saddle point for each realization of the 
Hubbard--Stratonovich field: to cancel terms linear in $W$.
Since we could not find the exact saddle point, such terms do 
appear, however, only in the second order in $\nabla K$. 
For $iS_{2}[\tilde Q,\nabla K]$ part of the action one obtains   
\begin{eqnarray}
iS_2^{(1)}[W,\nabla K] &=&
\frac{\pi\nu D}{2}  
 \nabla k_\alpha(\epsilon_1-\epsilon_2)
\Tr \left \{ \left[ 
  \gamma^{\alpha} \Lambda_{\epsilon_2} \gamma^{\beta } \Lambda_{\epsilon_3} -
  \Lambda_{\epsilon_1} \gamma^{\alpha} \Lambda_{\epsilon_2} \gamma^{\beta } 
\right] W \right\} \nabla k_\beta(\epsilon_2 - \epsilon_3) 
                                                              \nonumber \\
&=& \pi\nu D \Tr \left\{  
\nabla K^T(\epsilon_1-\epsilon_2)\left[  
M^{w}_{\epsilon_2}
w_{\epsilon_3\epsilon_1} +
M^{\overline w}_{\epsilon_1\epsilon_2\epsilon_3}
\overline w_{\epsilon_3\epsilon_1} \right]
\nabla K(\epsilon_2-\epsilon_3) \right\}  \, ,
                                                              \label{l8}
\end{eqnarray} 
where we have introduced two vertex matrices in the bosonic Keldysh space
\begin{equation} 
M^{w}_{\epsilon_2} =
\left( \begin{array}{cc}
 0     &  1 \\
-1     & -2F_{\epsilon_2}
\end{array} \right) \,; \,\,\,\, 
M^{\overline w}_{\epsilon_1\epsilon_2\epsilon_3} = 
\left( \begin{array}{cc}
2F_{\epsilon_2}-F_{\epsilon_1}-F_{\epsilon_3}      &  
1+F_{\epsilon_1}F_{\epsilon_3}-2F_{\epsilon_2}F_{\epsilon_3} \\
-1-F_{\epsilon_1}F_{\epsilon_3}+2F_{\epsilon_2}F_{\epsilon_1}\,\,\, & 
F_{\epsilon_1}+F_{\epsilon_3}-2F_{\epsilon_1}F_{\epsilon_2}F_{\epsilon_3}
\end{array} \right) \, .
                                                              \label{l9} 
\end{equation}
The fact that $(M^{w}_{\epsilon_2})^{1,1} = 0$ 
is a manifestation of the normalization condition, $Z=1$. Indeed, this
matrix element connects only the classical components of $W$ and $K$
fields, which alone can not change the normalization. Being averaged
over fluctuations of $\nabla K$ with the action  
Eq.~(\ref{u21}),  $iS_2^{(1)}[W,\nabla K]$ gives 
\begin{equation} 
\langle iS_2^{(1)}[W,\nabla K] \rangle_{\nabla K} =
2 \pi i \overline w_{\epsilon \epsilon}
 \left[ 
(F_{\epsilon+\omega} -F_{\epsilon}){\cal D}^{K}(\omega) - 
(1- F_{\epsilon+\omega} F_{\epsilon})
({\cal D}^{R}(\omega) - {\cal D}^{A}(\omega) )
\right] \, .
                                                              \label{l10} 
\end{equation} 
There is no term proportional to the classic component, $w$. In
equilibrium the r.h.s. of Eq.~(\ref{l10}) is obviously zero. Out of
equilibrium, it is this term  which is responsible for the standard
collision integral, see section \ref{sec.kinetic}. 
As we shall see in the next section, fluctuations described by 
$iS_2^{(1)}[W,\nabla K]$ are responsible for the Altshuler--Aronov
corrections to conductivity. 
For completeness we write also the second order expansion of 
$iS_1[W,\nabla K]$ 
\begin{eqnarray}
&&iS_1^{(2)}[W,\nabla K]=
i\nu D \Big[   
\nabla k_1(\epsilon_1-\epsilon_2)
(\nabla w_{\epsilon_2\epsilon_3} \overline  w_{\epsilon_3\epsilon_1}- 
\overline  w_{\epsilon_2\epsilon_3} \nabla w_{\epsilon_3\epsilon_1}) +
                                                       \label{l11} \\
&& \nabla k_2(\epsilon_1-\epsilon_2)
(-F_{\epsilon_2} \nabla \overline w_{\epsilon_2\epsilon_3}w_{\epsilon_3\epsilon_1}- 
w_{\epsilon_2\epsilon_3} \nabla \overline w_{\epsilon_3\epsilon_1}F_{\epsilon_1}+
B_{\epsilon_1-\epsilon_2}\nabla
(w_{\epsilon_2\epsilon_3} \overline  w_{\epsilon_3\epsilon_1}- 
\overline  w_{\epsilon_2\epsilon_3} w_{\epsilon_3\epsilon_1})) \Big]\, .
                                                      \nonumber
\end{eqnarray}

\subsection{Altshuler--Aronov corrections}
\label{subsec.AA}

\begin{figure}
\vglue 0cm
\hspace{0.01\hsize}
\epsfxsize=0.8\hsize
\epsffile{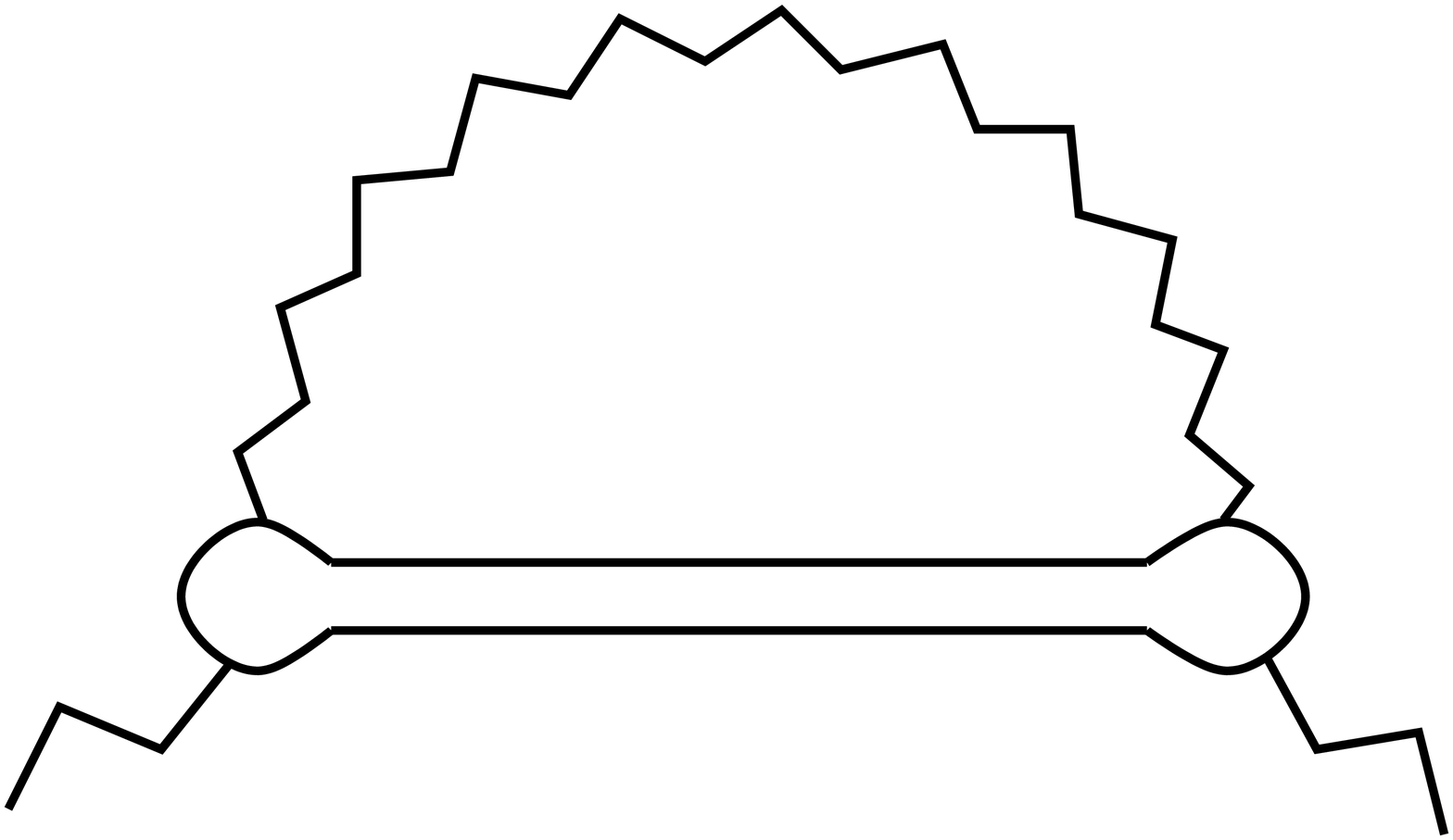}
\refstepcounter{figure} \label{zigzag:fig}
{\\ \small FIG.\ \ref{zigzag:fig}
Lowest order self--energy diagram for  
$\langle \nabla K \nabla K^T\rangle$ propagator. 
The zigzag lines represent the bare 
$\langle \nabla K \nabla K^T\rangle$ propagators, the parallel
solid lines denote the $\langle W W \rangle$ propagator and the open 
circles with two zigzag and two straight lines emanating from them represents 
the $\nabla K^T W  \nabla K$ vertices.  \par}
\end{figure}

Restricting oneself to the lowest non--vanishing terms in the expansion 
over $W$, Eqs.~(\ref{l5}) and (\ref{l8}), one obtains a Gaussian theory
with respect to the $W$ fluctuations. After integrating out these fluctuations
employing Eq.~(\ref{l5a}),
one ends up with the action for the $\nabla K$ field only
\begin{eqnarray}
&&iS[\nabla K]=
-i\nu D 
\Tr\left\{ \nabla K^T_{-\omega}(r) {\cal D}^{-1}(r-r',\omega) 
\nabla K_{\omega}(r') \right\} - 
                                                        \label{l12}\\      
&& 2\pi\nu D^2
\Tr\left\{ \nabla K^T_{\epsilon_1-\epsilon_2}(r)
M^{\overline w}_{\epsilon_1\epsilon_2\epsilon_3}
\nabla K_{\epsilon_2-\epsilon_3}(r) \right\}
D^{A}(r-r',\epsilon_3-\epsilon_1) 
\Tr\left\{ \nabla K^T_{\epsilon_3-\epsilon_4}(r')
M^{w}_{\epsilon_4}
\nabla K_{\epsilon_4-\epsilon_1}(r') \right\} \, ,
                                                          \nonumber
\end{eqnarray} 
This way the $(\nabla K)^4$ effective vertex is generated. 
Perturbatively the $(\nabla K)^4$ interaction term 
may be treated by pairing two fields, say 
$\nabla K^T \langle \nabla K \nabla K^T \rangle \nabla K$. This
results in  a renormalization of the bare correlator, ${\cal D}^{-1}$. 
Only pairing of $\nabla K$ fields in different spatial points leads to 
non--vanishing corrections, see Fig. \ref{zigzag:fig}. 
There are four different ways one can pair 
$\langle \nabla k_\alpha(r) \nabla k_\beta(r') \rangle$. Taking into
account all these four possibilities and integrating over an 
intermediate energy one obtains  
correction for e.g. retarded component of the 
$(\langle \nabla K \nabla K^T\rangle)^{-1}$  correlator
\begin{equation} 
[\delta {\cal D}^{-1}(q, \omega)]^{R} =
-\frac{4}{d\nu} \sum\limits_{q',\omega'} 
\Big[
D^{R}(q+q',2\omega+\omega') [{\cal D}(q', \omega+\omega')]^{R}
- D^{R}(q+q',\omega+\omega') [{\cal D}(q', \omega')]^{R}
\Big] \omega' B_{\omega'} \, ,
                                                              \label{l13} 
\end{equation}
\bottom{-2.7cm}
\narrowtext
\noindent  
where $B_\omega$ is defined by Eq.~(\ref{q18}); $d$ is  dimensionality. 
Obviously, the correction preserves the retarded   character
of the corresponding component. In equilibrium, the correction to the 
Keldysh component obeys the fluctuation--dissipation relation 
\begin{equation} 
[\delta {\cal D}^{-1}]^{K} =
\coth{\omega \over 2T} \left[ 
[\delta {\cal D}^{-1}]^{R}- 
[\delta {\cal D}^{-1}]^{A}
\right]  \, .
                                                              \label{l14} 
\end{equation} 
Employing Eqs.~(\ref{u4}), (\ref{u19}) one obtains for the correction to 
the $q=0$ conductivity
\begin{eqnarray} 
\delta \sigma(\omega) = && -i {4 e^2 D \over d \omega}
\sum\limits_{q',\omega'} 
D^{R}(q',\omega+\omega') [{\cal D}(q', \omega')]^{R}\times 
                                                            \label{l15}\\
&&\left[ 
(\omega'-\omega) B_{\omega'-\omega}  -\omega' B_{\omega'}
\right]  \, .
                                                            \nonumber
\end{eqnarray}
In the low frequency limit this reduces to the familiar expression 
\cite{Altshuler85} 
\begin{equation} 
\delta \sigma  = i {2 \sigma_d \over \pi d \nu}
\int\limits_{-\infty}^{\infty} \!\! d\omega   
{\partial \over \partial \omega}
\left(  \omega\coth{\omega \over 2T} \right)
\sum\limits_{q}
\frac{1}{(Dq^2 - i\omega)^2}    \, ,
                                                              \label{l16} 
\end{equation} 
where $\sigma_d = e^2 \nu D$. 
\begin{figure}
\vglue 0cm
\hspace{0.01\hsize}
\epsfxsize=0.9\hsize
\epsffile{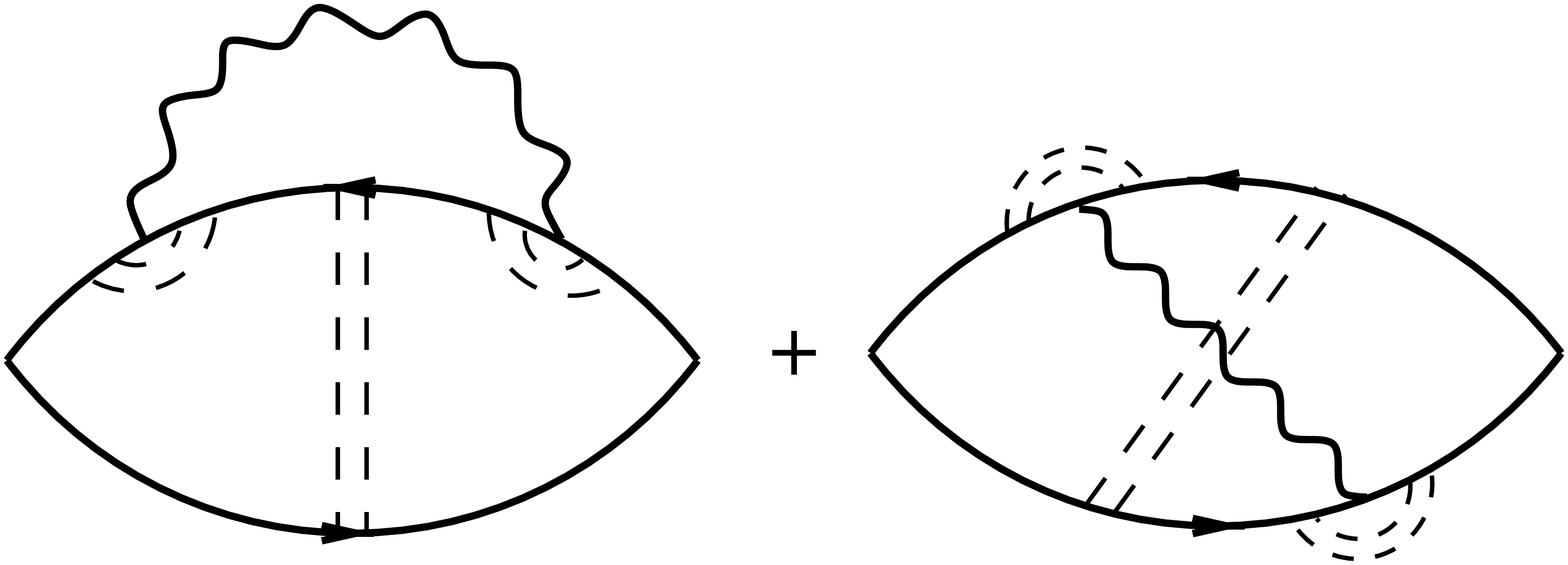}
\refstepcounter{figure} \label{AA:fig}
{\small FIG.\ \ref{AA:fig}
Lowest order diagrams for the interaction correction to
conductivity. Their sum is equivalent to the diagram 
in Fig.~\ref{zigzag:fig} in the present formalism. \par}
\end{figure}

\begin{figure}
\vglue 0cm
\hspace{0.01\hsize}
\epsfxsize=0.9\hsize
\epsffile{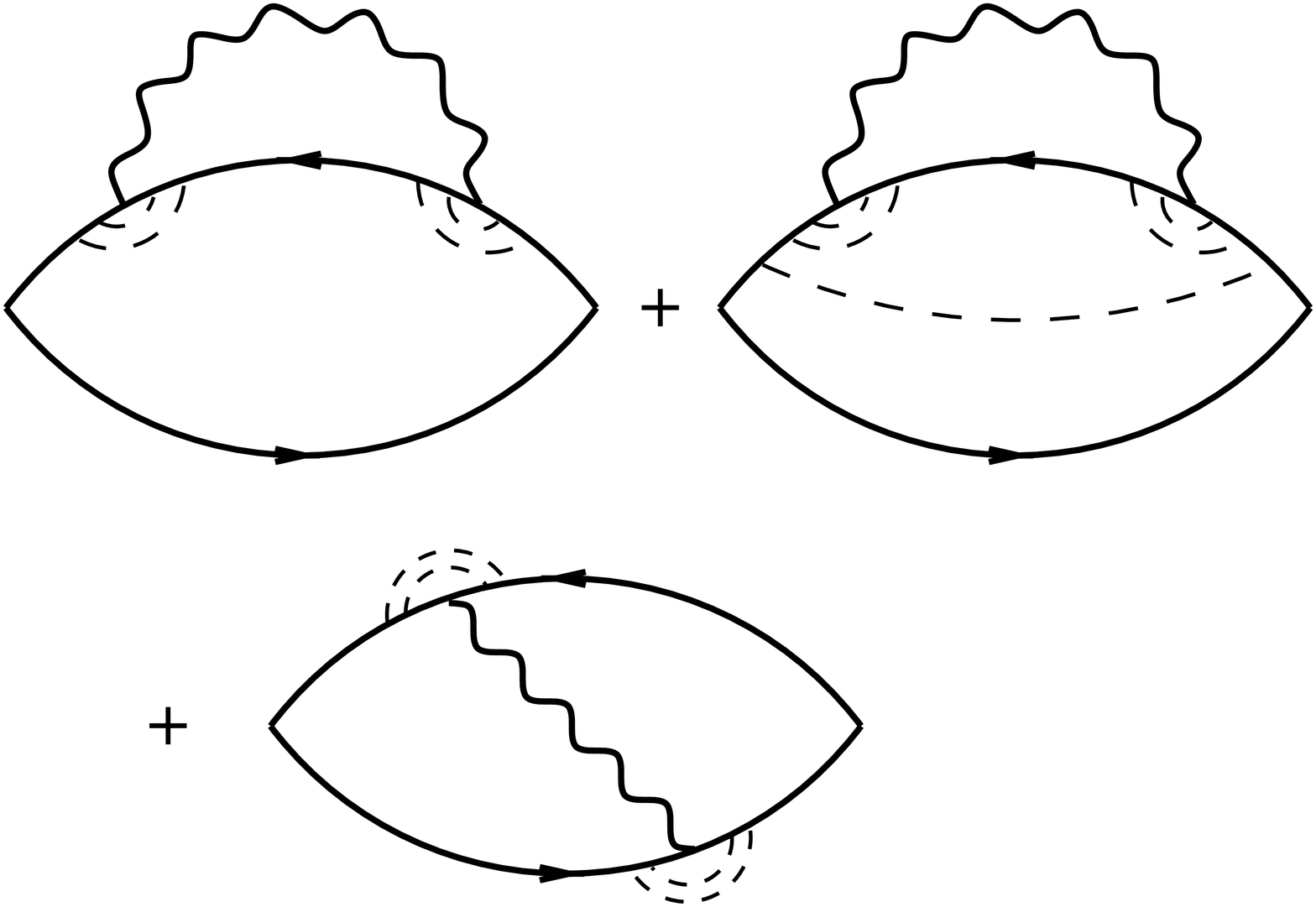}
\refstepcounter{figure} \label{cancel:fig}
{\small FIG.\ \ref{cancel:fig} 
Diagrams for the interaction corrections to conductivity
which add up to zero. These diagrams never appear in our formalism.
 \par}
\end{figure}

Note that this expression is given by the sum of diagrams drawn in
Fig.~\ref{AA:fig}.  The other diagrams which are presented in
Fig.~\ref{cancel:fig} add up to zero. They represent the purely phase
correction to the single particle Green function and therefore do not
enter the expression for the conductivity. In the present formalism 
these diagrams do not appear at all. 

In two dimensions  expression (\ref{l16}) leads to the logarithmically
divergent {\em negative} correction to the conductivity (or conductance) 
\begin{equation} 
\frac{\delta \sigma}{\sigma} = 
\frac{\delta g}{g} = {e^2 \over 2\pi^2 g} \ln T\tau^{el} \, , 
                                                              \label{l17} 
\end{equation} 
where the elastic mean free time, $\tau^{el}$, enters as an upper cutoff 
in the integral over frequency.

To handle this divergence one may try to setup a self--consistent mean--field
treatment of the $(\nabla K)^4$ non--linearity. To this end let us put 
$\langle \nabla K \nabla K^T \rangle$ propagator on Fig. \ref{zigzag:fig} to be 
a dressed one, $\tilde {\cal D}$, where  
$\tilde {\cal D}^{-1} = {\cal D}^{-1}+ \delta {\cal D}^{-1}$. Then 
Eq.~(\ref{l13}) may be rewritten as a closed non--linear equation for e.g. 
retarded  component of the propagator, 
$[\tilde {\cal D}(q.\omega)]^R$, 
\widetext
\top{-2.8cm} 
\begin{equation}
\left[  
Dq^2 - i\omega 
-\frac{4}{d \nu} \sum\limits_{q',\omega'} 
D^{R}(q+q',\omega+\omega') [\tilde {\cal D}(q',\omega')]^{R}
\left[(\omega'-\omega) B_{\omega'-\omega}-\omega' B_{\omega'} \right] 
\right]   [\tilde {\cal D}(q,\omega)]^{R}=1  \, .
                                                              \label{l18} 
\end{equation}
\bottom{-2.7cm}
\narrowtext
\noindent  
The frequency dependent conductivity is then given by 
\begin{equation}
\sigma (\omega)=e^2 \nu D \, 
{ ([\tilde {\cal D}(q=0,\omega)]^{R})^{-1} \over -i\omega}\, .
                                                              \label{l19} 
\end{equation}
One may easily check that in the one--loop approximation there are no 
other corrections to  conductivity. Indeed, the possible corrections like 
$\langle \left( iS_1^{(2)}[W,\nabla K] \right)^2 \rangle_{W} =0$ and 
$\langle iS_2^{(2)}[W,\nabla K]  \rangle_{W} =0$.
They vanish since they include 
energy integration of purely retarded or advanced functions. Being expanded
to higher orders in $W$ these terms yield weak--localization corrections.

\section{Kinetic Equation}
\label{sec.kinetic}

The aim of this section is to demonstrate how the kinetic equation 
for the distribution function $F$ appears naturally in the framework 
of the Keldysh formulation. The kinetic equation is 
nothing but the saddle point equation for the effective action on the 
$\tilde Q$--matrix \cite{MuzykantskyKhmelnitskii}. In the case of
interacting electrons it is obtained by integrating out 
the $K$ (or equivalently $\Phi$) degrees of freedom. Consider the partition
function, Eq.~(\ref{r6}), with the action, $S(\tilde Q, \Phi)$, given 
by Eq.~(\ref{q8}). Let us perform the $\Phi$ integration first. As a
result we obtain  for the average partition function 
\begin{mathletters}
\label{k1,2}
\begin{eqnarray} 
&&\langle Z \rangle = 
\int {\cal D} \tilde Q\,  e^{iS_{eff}[\tilde Q]}\, ,
                                                              \label{k1}\\
&&iS_{eff}[\tilde Q] =
\ln \int {\cal D} \Phi e^{i\,\Tr \{ \Phi^T V_0^{-1}\sigma_1 \Phi \} +
i S[\tilde Q, \Phi]}\, .
                                                              \label{k2}
\end{eqnarray}
\end{mathletters}    
Since the the action, $S(\tilde Q, \Phi)$, Eq.~(\ref{q8}), is
quadratic in $\Phi$ (given the linear relation between $K$ and $\Phi$)
the integration in the last expression can be carried out
explicitly. We find it more convenient, however, to proceed with the
expression (\ref{k2}). To obtain a non--trivial kinetic
theory one may assume the presence of classical external fields, like
e.g. scalar or vector potentials. These fields can be introduced in
the action  Eq.~(\ref{q8}) in the way it was done in section
\ref{subsec.sources}.  

We shall look now for the saddle point equation for $\tilde Q$  
\begin{equation}
\frac{\delta S_{eff}[\tilde Q]}{\delta \tilde Q} = 0 \, ,
                                                              \label{k3}
\end{equation}
obtained under the condition $\tilde Q^2 =1$. Let us reiterate again 
the logic of the entire procedure. After averaging over disorder and 
introducing the $\tilde Q$--matrix, we found that the low--energy
degrees of freedom are described by the $\tilde Q$--matrices given by
Eq.~(\ref{p9a}) with $\tilde Q^2 =1$. We then restrict ourselves to
this massless 
manifold and look for a realization of $\tilde Q$ which extremize the 
effective action. The latter is obtained by integrating out the photon
fields originating from e--e interactions.  
Without any external fields (and/or non--trivial boundary conditions)
such an extremal  $\tilde Q$ is simply given by $\Lambda$,
Eq.~(\ref{p6}), with the  
equilibrium $F$ function, Eq.~(\ref{p5}).  If  external fields   
(and/or non--trivial boundary conditions) are present, the stationary 
$\tilde Q$ may deviate from $\Lambda$, still being on the massless
manifold, $\tilde Q^2 =1$. The stationary point is to be found by
solving Eq.~(\ref{k3}), which turns out to be precisely the kinetic
equation with the collision integral term. 

Before proceeding along these lines, let us comment on the relation between 
the phase $K$, introduced in section \ref{subsec:sp} and the 
Hubbard--Stratonovich field, $\Phi$. The procedure of section \ref{subsec:sp}
was based on the property of the equilibrium distribution described by 
Eq.~(\ref{p20}). We need to generalize it for non--equilibrium  
situations. To this end we note that the equation for the quantum component, 
$k_2(r,t)$, Eq.~(\ref{p18}), does not contain distribution function and 
remains valid for a non--equilibrium case. The equation for the 
classical component $k_1(r,t)$, Eq.~(\ref{p19}), can not be satisfied 
identically out of equilibrium. Thus the choice of $k_1(r,t)$ allows for a 
certain arbitrariness. 
However, as we shall see below, this arbitrariness {\em does not} affect 
the form of the kinetic (saddle point) equation. It would manifest itself 
in a calculation of fluctuation corrections (cf. section \ref{sec.fluct} )
to the non--equilibrium saddle point result. We shall not attempt this task 
here. For our purposes it is sufficient to keep the definition of $K(r,t)$ 
given by Eq.~(\ref{p21}) (or Eq.~(\ref{u5}) if external fields are present). 
The equilibrium bosonic distribution function, used in the definition of the 
Keldysh component of  the propagator ${\cal D}(q,\omega)$, Eq.~(\ref{p23}b), 
does not show up in the kinetic equation.

Employing Eq.~(\ref{k2}), we rewrite the saddle point equation
(\ref{k3})  as 
\begin{equation}
\left\langle
\frac{\delta S[\tilde Q,\Phi]}{\delta \tilde Q} 
\right\rangle_{\Phi}  = 0 \, ,  
                                                            \label{k4}
\end{equation}
where
\begin{equation}
\big\langle
\ldots  
\big\rangle_{\Phi}  = 
\frac{\int {\cal D} \Phi e^{i\, \Tr\{ \Phi^T V_0^{-1}\sigma_1 \Phi \}  +
i S[\tilde{\underline Q}, \Phi]} \ldots }
{\int {\cal D} \Phi e^{i\, \Tr\{ \Phi^T V_0^{-1}\sigma_1 \Phi \} +
i S[\tilde{\underline Q}, \Phi]}  }
 \, .  
                                                            \label{k5}
\end{equation}
Here $\tilde{\underline Q}$ is a self--consistent  saddle point solution 
of  Eq.~(\ref{k4}). Performing variation of the action $S(\tilde Q, \Phi)$ 
given by Eq.~(\ref{q8}) under the condition $\tilde Q^2=1$, one obtains 
\begin{equation}
\left\langle
D \partial_r ( \tilde{\underline Q} \partial_r\tilde{\underline Q} ) +
i [(\epsilon +(\phi_\alpha+i\omega k_{\alpha})\gamma^{\alpha}), 
\tilde{\underline Q} ] 
\right\rangle_{\Phi}  = 0 \, , 
                                                            \label{k6}
\end{equation}
where $\tilde{\underline Q}^2=1$. This equation is analogous to the
kinetic equation in the semiclassical theory of disordered superconductors
\cite{Usadel,Larkin}. We have derived it 
here for the case of a normal interacting metal.   

We shall seek the solution of  Eq.~(\ref{k6}) in the classical form, e.g. 
obeying the condition $\tilde{\underline Q}_{21} = 0$. A non--zero quantum 
component at the saddle point would violate
causality. Provided $\tilde{\underline Q}_{21} = 0$ and
$\tilde{\underline Q}^2=1$ are satisfied  
the saddle point solution assumes the form 
\begin{equation} 
\tilde{\underline Q}_{\epsilon,\epsilon'} = 
\left( \begin{array}{cc}
\openone_{\epsilon}^R \delta(\epsilon-\epsilon') & 
2F_{\epsilon,\epsilon'}(r) \\ 
0  & -\openone_{\epsilon}^A \delta(\epsilon-\epsilon') 
\end{array} \right)\, ,
                                                              \label{k7}
\end{equation}
where $F_{\epsilon,\epsilon'}(r)$ is a non--stationary distribution function. 
Assuming that the saddle point has the form given by  Eq.~(\ref{k7}), 
one can easily check that the exponent in the $\Phi$ averaging, 
Eq.~(\ref{k5}), does not contain linear terms in $\Phi$ (or $\nabla 
K$). Indeed, the terms proportional to  
$\nabla k_1$ vanish identically, which is a manifestation of the 
normalization condition, $Z=1$. From another hand, terms 
proportional to $\nabla k_2$ are 
reduced to the full gradient (the fact that there is no  ambiguity in the 
choice of $k_2$ is important here) and thus also vanish upon the spatial 
integration.  As a result the  terms linear in $\Phi$ (or $\nabla K$) in 
the saddle point equation (\ref{k6}) do not survive  the $\Phi$ 
integration. Therefore Eq.~(\ref{k6}) may be reduced to 
\begin{equation}
D \nabla_r ( \tilde{\underline Q} \nabla_r\tilde{\underline Q} ) +
i [\epsilon, \tilde{\underline Q}] = 
D \langle
[\nabla k_{\alpha}\gamma^{\alpha}\tilde{\underline Q}
\nabla k_{\beta}\gamma^{\beta}, \tilde{\underline Q}]
\rangle_{\Phi}\, .
                                                              \label{k8}
\end{equation}
The r.h.s. of this  equation contains the collision integral term 
along with the collisionless renormalization of the kinetic part. 
To evaluate it one needs to know the propagator 
$\langle \nabla k_{\alpha}(r,t)\nabla k_{\beta}(r',t')\rangle_{\Phi}$ 
at $r=r'$, averaged over the non--equilibrium action, Eq.~(\ref{k5}). 
To follow the same notations as for the equilibrium case we shall denote 
this propagator as 
\begin{equation}
\langle \nabla k_{\alpha}(r,t)\nabla k_{\beta}(r',t')\rangle_{\Phi}=
- \frac{i}{2\nu D} \, {\cal D}^{\alpha\beta}_{t,t'}(r,r')\, .
                                                              \label{k9}
\end{equation}
The form of the saddle point, $ \tilde{\underline Q}$ given by Eq.~(\ref{k7})
guarantees that ${\cal D}_{t,t'}$ has the standard retarded/advanced structure 
of a  Keldysh propagator. Employing Eq.~(\ref{p7}), one finds that the only 
non--zero matrix component of  Eq.~(\ref{k8}) is its Keldysh $(1,2)$ component.
The corresponding equation for the distribution function $F_{t,t'}(r)$ 
takes the following form  
\widetext
\top{-2.8cm}
\begin{equation}
D \nabla^2_r F_{t,t'} - (\partial_{t} + \partial_{t'}) F_{t,t'} 
 = {i\over \nu} 
\left[  
F_{t,t'} \left( 
{\cal D}^K_{t,t'} - {1\over 2}[{\cal D}^K_{t,t}+{\cal D}^K_{t',t'}]
\right) + 
({\cal D}^R_{t,t_1} - {\cal D}^A_{t_1,t'})
(\delta_{t,t_1}\delta_{t_1,t'} - F_{t,t_1}F_{t_1,t'})
\right]   \, .
                                               \label{k10}     
\end{equation}
Here all  $F$ functions and propagators ${\cal D}$ are to be taken 
at the same spatial point;  integration over $t_1$ is assumed in the last term 
on the r.h.s. Note that the l.h.s. of this equation is a linear diffusion 
operator acting on $F_{t,t'}(r)$. The subsequent calculations are
significantly simplified by passing to the Wigner representation, 
\begin{equation}
F_{\epsilon}(r,\tau) = 
\int\!\! \int dt dt' F_{t,t'}(r) e^{i\epsilon (t-t')}
\delta \left( \tau - {t+t'\over 2} \right) \, . 
                                                              \label{k11}
\end{equation}
Furthermore we shall assume that $F_{\epsilon}(r,\tau)$ is a slow function of 
$\tau$ on the scale $1/T$ (or any other inverse characteristic scale of energy, 
$\epsilon$). With this assumption Eq.~(\ref{k10}) may be rewritten as 
\begin{eqnarray}
&&D \nabla^2_r F_{\epsilon}(\tau) - \partial_{\tau}F_{\epsilon}(\tau) -
\partial_{\tau}{\cal R}_{\epsilon}(\tau)\partial_{\epsilon}F_{\epsilon}(\tau)+
\partial_{\epsilon}{\cal R}_{\epsilon}(\tau)\partial_{\tau}F_{\epsilon}(\tau) 
                                                              \label{k12}\\
&&={i\over \nu} 
\sum\limits_{\omega}
\Big[ 
{\cal D}^K_{\omega}(\tau) 
(  F_{\epsilon-\omega}(\tau) - F_{\epsilon}(\tau)  ) + 
(  {\cal D}^R_{\omega}(\tau) - {\cal D}^A_{\omega}(\tau)  ) 
(  1 - F_{\epsilon-\omega}(\tau)  F_{\epsilon}(\tau)  )
\Big]   \, ,
                                                         \nonumber 
\end{eqnarray}
where 
\begin{equation}
{\cal R}_{\epsilon}(r,\tau) = {1\over 2\nu} 
\sum\limits_{\omega}
\Big[ {\cal D}^R_{\omega}(r,r,\tau) + {\cal D}^A_{\omega}(r,r,\tau)  \Big]
F_{\epsilon-\omega}(r,\tau)\, . 
                                                              \label{k13}
\end{equation}
The r.h.s. of Eq.~(\ref{k12}) represents the collision integral,
cf. Eq.~(\ref{l10}).    If equilibrium 
relation, Eq.~(\ref{p23}b), between Keldysh and retarded and advanced 
components of ${\cal D}$ holds, then the equilibrium distribution function,  
Eq.~(\ref{p5}), nullifies the collision integral. Therefore Eq.~(\ref{k12}) 
is satisfied in the thermal equilibrium. This, in fact, 
provides  justification for our previous use of $\Lambda$ with the
equilibrium $F$--function as the saddle point. Indeed, without
interactions (and hence without collision integral) any stationary
function $F_\epsilon$ satisfies the saddle point equation. It is the
relaxation processes due to e--e interactions that render the
equilibrium solution unique. The terms which contain real part of  
the self--energy, ${\cal R}_{\epsilon}(r,\tau)$, lead to a collisionless 
renormalization of the kinetic part, see section \ref{subsec.colless}. 

To proceed further we need an explicit form of the non--equilibrium 
propagator, ${\cal D}_{\omega}(r,r,\tau)$. We shall evaluate it in the 
universal limit of strong interactions, $V_0^{-1}\rightarrow 0$. Substituting the 
saddle point  $ \tilde{\underline Q}$ given by Eq.~(\ref{k7}) into the 
action $S[\tilde{\underline Q},\Phi]$, Eq.~(\ref{q8}), and performing the 
Gaussian integration one finds (cf.  Eq.~(\ref{q16}) )
\begin{equation}
{\cal D}^{\alpha\beta}_{\omega}(r,r',\tau) = 
\left[
-D\nabla^2_r \sigma^{\alpha\beta}_1 + \delta_{r,r'}  
\frac{i\pi}{2} \sum\limits_{\epsilon} 
\Tr \left\{
\gamma^{\alpha}\tilde{\underline Q}_{\epsilon+{\omega \over 2}}(r,\tau)
\gamma^{\beta} \tilde{\underline Q}_{\epsilon-{\omega \over 2}}(r,\tau) -
\gamma^{\alpha}\gamma^{\beta}
\right\} 
\right]^{-1}  \, . 
                                                              \label{k14}
\end{equation}
The term with $\nabla^2_r$ originates from the term 
$|\Phi+i\omega K|^2\sigma_1$ in the action  Eq.~(\ref{q8}). 
(It is easy to check that the ambiguity in the choice of $k_1$, 
mentioned above, disappears upon the calculation of this term 
by the symmetry reason.)
The local in space term in Eq.~(\ref{k14}) originates 
from $D\Tr (\partial_r \tilde Q)^2$.  Assuming that 
any distortion of the equilibrium distribution is limited to a
vicinity  of the Fermi energy, e.g. 
$F_{\epsilon\rightarrow \pm \infty}(r,\tau)\rightarrow \pm 1$, 
one finds   
\begin{equation}
{\cal D}_{\omega}(r,r',\tau) = 
\left( 
\begin{array}{cc}
0 & -D\nabla^2_r  + i\omega \delta_{r,r'}  \\ 
-D\nabla^2_r  -  i\omega \delta_{r,r'} & 
-2i\omega \delta_{r,r'}B_\omega(r,\tau)
\end{array}
\right)^{-1}  \, .
                                                              \label{k15}
\end{equation}
By definition, the non--equilibrium bosonic distribution function 
is given by  (cf.  Eq.~(\ref{q18}) )
\begin{equation}
B_\omega(r,\tau) = {1\over 2\omega }
\int\limits_{-\infty}^{\infty}\!\!  d\epsilon\,  
\big[ 1 - F_{\epsilon+{\omega \over 2}}(r,\tau)
F_{\epsilon-{\omega \over 2}}(r,\tau) \big]\, .
                                                              \label{k16}
\end{equation}
According to Eq.~(\ref{k15}) the retarded and advanced components of 
${\cal D}$ are not modified with respect to their equilibrium value, 
Eq.~(\ref{p23}a). As a result, 
${\cal D}^{R(A)}_{\omega}(r,r',\tau) = {\cal D}^{R(A)}_{\omega}(r - r')$ 
even in a non--equilibrium situation.  
Inverting the operator on the r.h.s. of  Eq.~(\ref{k15}), 
one finds for the Keldysh component at coinciding spatial points 
\begin{eqnarray}
&&{\cal D}^K_{\omega}(r,r,\tau) = 
2i \omega \!\!
\int \! d^d r'\, 
\Bigg[ 
{\cal D}^R_{\omega}(r-r')B_\omega(r',\tau){\cal D}^A_{\omega}(r'-r)
  + 
                                                         \label{k17}\\
&& \frac{1}{2i} \Big[
{\cal D}^R_{\omega}(r-r') \partial_\tau  B_\omega(r',\tau)
\partial_\omega {\cal D}^A_{\omega}(r'-r) - 
\partial_\omega {\cal D}^R_{\omega}(r-r') 
\partial_\tau  B_\omega(r',\tau) {\cal D}^A_{\omega}(r'-r) \Big]
\Bigg] 
                                                      \nonumber
\end{eqnarray}
>From now on we shall retain only the first term in this expression, 
which is dominant due to the assumed slowness of the temporal 
variations of $F_{\epsilon}(\tau)$. 
If in addition $B_\omega(r,\tau)$ changes slowly on the spatial scale 
$L_\omega = \sqrt{D/\omega}$, where $\omega\sim T$, the expression for 
the Keldysh component acquires  the quasi-equilibrium form 
\begin{equation}
{\cal D}^K_{\omega}(r,r,\tau) = 
B_\omega(r,\tau) \sum\limits_q 
[ {\cal D}^R_{\omega}(q) - {\cal D}^A_{\omega}(q)  ] \, .
                                                              \label{k18}
\end{equation}
One can calculate  gradient corrections to this expression, which  
lead to a non--local collision integral. Usually such corrections may be 
safely neglected. 
Finally in this  hydrodynamic regime the kinetic equation takes the form  
\begin{eqnarray}
&&D \nabla^2_r F_{\epsilon}(\tau) - 
[1 - \partial_{\epsilon}{\cal R}_{\epsilon}(\tau) ] 
\partial_{\tau}F_{\epsilon}(\tau) -
\partial_{\tau}{\cal R}_{\epsilon}(\tau)\partial_{\epsilon}F_{\epsilon}(\tau)  
                                                              \label{k19}\\
&& = -  \sum\limits_{\omega}
\left[ {2\over \nu} \Im \sum\limits_{q}{\cal D}^R_{\omega}(q) \right]   
\Big[ 
B_\omega(\tau)  (  F_{\epsilon-\omega}(\tau) - F_{\epsilon}(\tau)  ) + 
(  1 - F_{\epsilon-\omega}(\tau)  F_{\epsilon}(\tau)  )
\Big]   \, ,
                                                         \nonumber 
\end{eqnarray}
with 
\begin{equation}
{\cal R}_{\epsilon}(r,\tau) = {1\over \nu} 
\sum\limits_{\omega,q} 
\left[ \Re {\cal D}^R_{\omega}(q) \right]  
F_{\epsilon-\omega}(r,\tau)\,  
                                                              \label{k20}
\end{equation}
and ${\cal D}^R_{\omega}(q) =(Dq^2 - i\omega)^{-1}$.

\subsection{Collision integral and relaxation time}
\label{subsec.collision}

\begin{figure}
\vglue 0cm
\hspace{0.01\hsize}
\epsfxsize=0.9\hsize
\epsffile{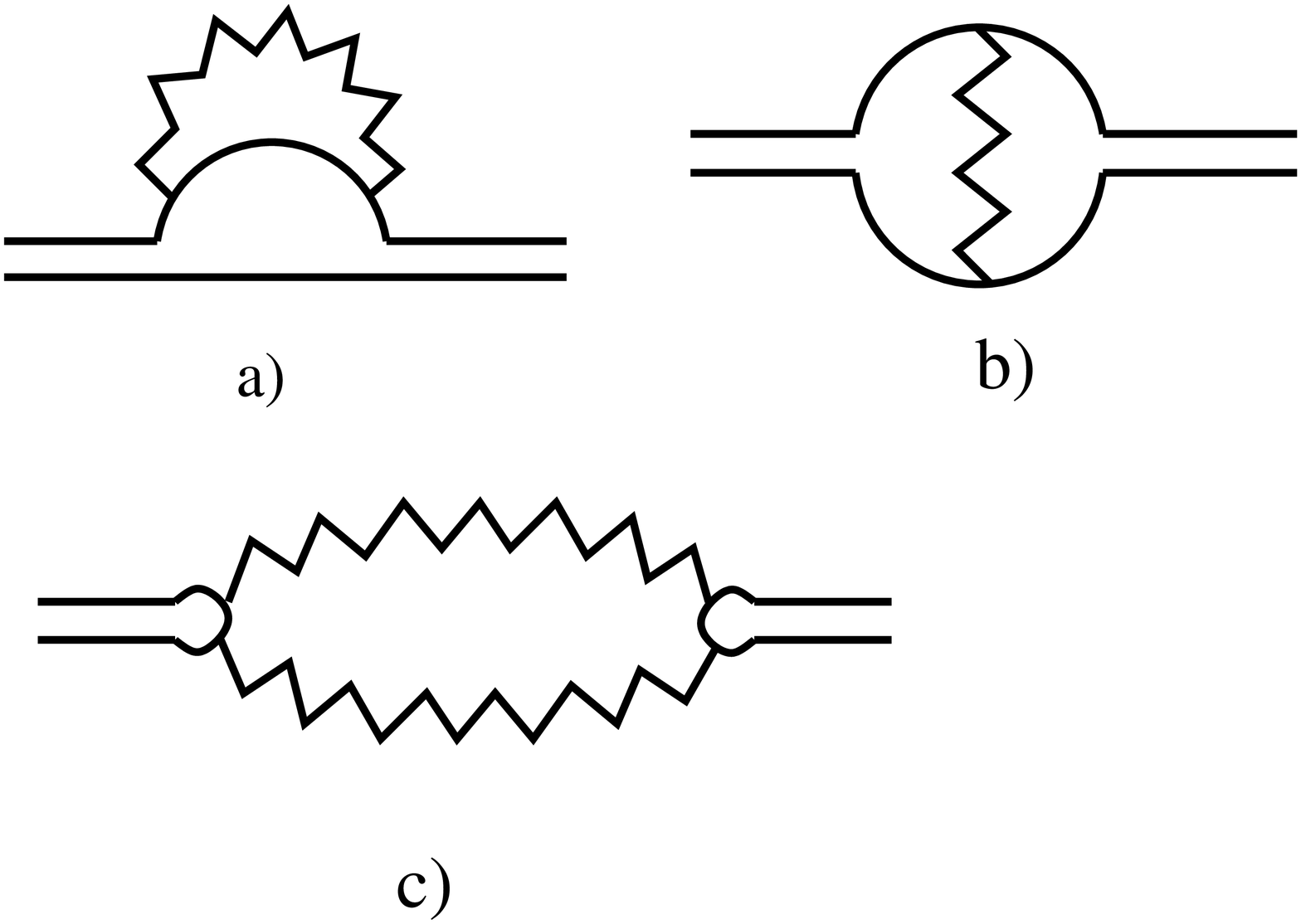}
\refstepcounter{figure} \label{reltime:fig}
{\small FIG.\ \ref{reltime:fig} 
Diagrammatic representation of the Gaussian part of the effective action,
$S_{eff}$, Eq.~(\ref{k2}). $a)$ and $b)$ represent ``out'' and ``in'' 
relaxation terms correspondingly; they originates from the first term 
in Eq.~(\ref{k23}), $\langle iS_2 \rangle_{\nabla K}$. 
Non--local term $c)$ arises from the second term in  Eq.~(\ref{k23}), 
$\langle iS_2 iS_2 \rangle_{\nabla K}$. 
 \par}
\end{figure}

Using the conventional fermion distribution function, 
$n_{\epsilon}(r,\tau) = (1-F_{\epsilon}(r,\tau))/2$, one can rewrite 
the collision integral in the usual form with  ``out'' and ``in'' 
relaxation terms. Indeed, employing Eqs.~(\ref{q17}), (\ref{q18}), 
one identically rewrites the r.h.s. of Eq.~(\ref{k19}) as 
\begin{equation}
 - \int\!\!\!\! \int\limits_{-\infty}^{\infty}
\frac{d\omega d\epsilon'}{\pi}\, 
\frac{4 \Im \sum\limits_{q}{\cal D}^R_{\omega}(q)}{\nu \omega}   
\Big[ 
n_{\epsilon}n_{\epsilon'-\omega}(1-n_{\epsilon'})(1-n_{\epsilon-\omega})- 
n_{\epsilon'}n_{\epsilon-\omega}(1-n_{\epsilon})(1-n_{\epsilon'-\omega})
\Big]   \, . 
                                                              \label{k21}
\end{equation}
This is precisely the collision term derived by Altshuler \cite{Altshuler78} 
and Altshuler and Aronov \cite{Altshuler79a} two decades ago. One can  
linearize this expression  around the equilibrium distribution 
by the substitution
\begin{equation}
F_\epsilon(r,\tau) = F_\epsilon^{eq} - 
w_\epsilon(r,\tau)/2 
                                                              \label{k22}
\end{equation}
and keeping linear terms in $w_\epsilon(r,\tau)$. This way one derives the 
familiar results for the relaxation rates  
\cite{Shmid74,Altshuler79a,Altshuler85}. 
We shall not repeat this procedure here. Instead we shall demonstrate
how this quantities may be extracted directly  
from the effective action. To this end we note that the kinetic equation 
(\ref{k12}) may be written as 
$(2\pi \nu)^{-1} \delta iS_{eff}/\delta \overline w_\epsilon(r,\tau) = 0$. 
(As usual, an observable is generated by differentiation with respect
to a quantum  component.) Thus the 
linearized version of the kinetic equation is just 
$-(\pi \nu)^{-1} \delta^2 iS_{eff}/\delta \overline w \delta w
|_{\tilde Q = \Lambda}$. According to  Eqs.~(\ref{k2}) and (\ref{k5}) 
\begin{equation}
\frac{\delta^2 iS_{eff}}{\delta \overline w \delta w}
  =  
\left\langle
\frac{\delta^2 iS[\tilde Q,\Phi] }{\delta \overline w \delta w} 
\right\rangle_{\Phi} + 
\left\langle
\frac{\delta iS[\tilde Q,\Phi] }{\delta \overline w }
\frac{\delta iS[\tilde Q,\Phi] }{\delta w }  
 \right\rangle_{\Phi} - 
\left\langle
\frac{\delta iS[\tilde Q,\Phi] }{\delta \overline w }  
 \right\rangle_{\Phi}
\left\langle
\frac{\delta iS[\tilde Q,\Phi] }{\delta w }
\right\rangle_{\Phi}
\, ,  
                                                            \label{k23}
\end{equation}
where all the variational derivatives are calculated at 
$\tilde Q = \tilde{\underline Q} = \Lambda$. The last term in this expression 
vanishes identically, since $\Lambda$ is obviously a solution of the 
kinetic equation (\ref{k4}). The first term originates from the expansion 
of $\langle iS_2[W,\nabla K] \rangle_{\nabla K}$, Eq.~(\ref{q13}), 
to the leading order in $w$ and  $\overline w$. After a little algebra one 
obtains (there are no terms with $\overline w  \overline w $ in equilibrium) 
\begin{equation}
\langle iS_2^{(2)}[W,\nabla K] \rangle_{\nabla K}= 
\frac{i\pi}{2} 
(\overline w_{\epsilon_+ -\omega,\epsilon_- -\omega} - 
\overline w_{\epsilon_+,\epsilon_-})
\Big[
{\cal D}^R_\omega (B_\omega^{eq} + F_{\epsilon_- -\omega}^{eq}) - 
{\cal D}^A_\omega (B_\omega^{eq} + F_{\epsilon_+ -\omega}^{eq})
\Big]
w_{\epsilon_-,\epsilon_+} \, , 
                                                            \label{k24}
\end{equation}
where $\epsilon_{\pm} = \epsilon \pm \Omega/ 2$. 
Equivalently this expression 
can be obtained by variation of the $F_\epsilon(r,\tau)$ functions in  
Eq.~(\ref{k10}). The terms with $\overline w_{\epsilon_+,\epsilon_-}$
and  $\overline w_{\epsilon_+-\omega,\epsilon_- -\omega} $ represent
``out'' and ``in'' relaxation processes respectively. Their
condensed diagrammatic representation is given in 
Fig. \ref{reltime:fig}a,b. The 
full set of corresponding original diagrams may be found e.g. in
Ref.~\cite{Blanter96}. Restricting ourselves to the diagonal  
fluctuations, $\Omega= 0$, only, we  obtain  for the 
``out'' relaxation rate 
\begin{equation}
\frac{1}{\tau_{out}(\epsilon,T) } = 
\sum\limits_{\omega q}  
\left[ {2\over \nu} \Im {\cal D}^R_{\omega}(q) \right]   
\Big[ 
\coth  \frac{\omega}{2T} + \tanh  \frac{\epsilon - \omega}{2T}
\Big] \, .
                                                        \label{k25}
\end{equation} 
At $T=0$ in two dimensions this leads to the familiar result
\cite{Altshuler85}   
\begin{equation}
\frac{1}{\tau^{d=2}_{out}(\epsilon) } = 
\frac{|\epsilon|}{4\pi g} \, ,
                                                        \label{k26}
\end{equation} 
where $g=\nu D$. 
Expanding the r.h.s. of Eq.~(\ref{k24}) in a small $\Omega$, 
one can also recover the collisionless terms in the l.h.s. of
Eq.~(\ref{k19}).

Finally we concentrate on the second term in the r.h.s. of
Eq.~(\ref{k23}). This term corresponds to the variation of ${\cal
D}^K_{t,t'}$ in Eq.~(\ref{k10}) 
over a deviation from its equilibrium value 
(or equivalently variation of $B_\omega(r,\tau)$ 
in Eq.~(\ref{k19}) ). Its condensed diagrammatic representation is
depicted on   Fig. \ref{reltime:fig}c. 
As has already been mentioned above, this term is generally 
spatially non--local. We take here only its local part. Technically 
it originates from a connected part of 
${1\over 2}\langle S_2^{(1)}[W,\nabla K] S_2^{(1)}[W,\nabla
K]\rangle_{\nabla K}$,  where 
$iS_2^{(1)}$ is given by  Eq.~(\ref{l8}). Performing averaging over 
$\nabla K$ and omitting cumbersome $\overline w\overline w$ terms, 
one obtains 
\begin{eqnarray}
\langle i S_2^{(1)} i S_2^{(1)} \rangle_{\nabla K} &=&
-\frac{\pi^2}{2}
\sum\limits_{\epsilon\epsilon'\omega \Omega q}
\overline w_{\epsilon_+,\epsilon_-} 
\Tr \left\{
[M^{w}_{\epsilon'+\omega} + (M^{w}_{\epsilon'-\omega})^T]
{\cal D}_{\omega +{\Omega\over 2}} 
(M^{\overline w}_{\epsilon_-, \epsilon-\omega, \epsilon_+})^T
{\cal D}_{\omega -{\Omega\over 2}} 
\right\} 
w_{\epsilon_-', \epsilon_+' }  
                                                 \label{k27} \\
&=& \pi^2  
\overline w_{\epsilon_+,\epsilon_-} 
\Big[ 
(2F_{\epsilon-\omega}^{eq} - F_{\epsilon_-}^{eq}  - F_{\epsilon_+}^{eq})
(F_{\epsilon'+ \omega}^{eq} + F_{\epsilon'-\omega}^{eq})
{\cal D}^R_{\omega -{\Omega\over 2}}
{\cal D}^A_{\omega +{\Omega\over 2}} 
\Big]
w_{\epsilon_-', \epsilon_+' } 
\, ,
                                                   \nonumber 
\end{eqnarray}
\bottom{-2.7cm}
\narrowtext
\noindent
For $\Omega=0$ this 
expression coincide with the variation of $B_\omega(r,\tau)$ over $w$ in 
Eq.~(\ref{k19}). Expanding to the first order in $\Omega$, one obtains 
the correction terms written in Eq.~(\ref{k17}).    
Eqs. (\ref{k23}), (\ref{k24}) and (\ref{k27}) along with 
Eq.~(\ref{l5}) complete calculations of 
$\delta^2 iS_{eff}/\delta \overline w \delta w$ on the mean--field level. 
Let us notice for completeness that 
$\delta^2 iS_{eff}/\delta w \delta w = 0$, which is a manifestation 
of the normalization condition. From another hand, 
$\delta^2 iS_{eff}/\delta \overline w \delta \overline w \neq 0$, originates 
solely from the second term on the r.h.s. of Eq. (\ref{k23}), cf. 
Fig. \ref{reltime:fig}c.

\subsection{Collisionless terms}
\label{subsec.colless}

Finally we briefly discuss the physics of the collisionless terms. 
Collisionless terms originate from the real part of the 
selfenergy, ${\cal R}_\epsilon(r,t)$, and thus appear already in the
first order in the bare interaction (unlike the collision integral,
which arises only in the second order). For the screened Coulomb
interaction one obtains from  
Eq. (\ref{k20})
\begin{equation}
{\cal R}_{\epsilon}(r,\tau) = 
\int\limits_{-\infty}^{\infty} \!\! 
\frac{d \omega}{2\pi} 
F_{\epsilon-\omega}(r,\tau)
\int\! (d^dq)
{1\over \nu} 
\frac{ Dq^2}{(Dq^2)^2 + \omega^2 } \,  .
                                                              \label{k28}
\end{equation}
In two dimensions this leads to the following  logarithmic expression
\begin{equation}
{\cal R}_{\epsilon}(r,\tau) = 
- \frac{1}{4\pi g} 
\int\limits_{-1/\tau^{el}}^{1/\tau^{el}} \!\!\!\! 
\frac {d \omega}{2\pi}
\ln (\tau^{el}  |\omega|)  
F_{\epsilon-\omega}(r,\tau) \,  ,
                                                              \label{k29}
\end{equation}
where we have used the superscript for the elastic mean free time,
$\tau^{el}$, to avoid 
confusion with a physical time, $\tau$.   If one linearize the kinetic
equation  (\ref{k19}) around the equilibrium distribution, its
l.h.s. acquires the form  
\begin{eqnarray}
D \nabla^2_r F_{\epsilon}(\tau) && - 
\left[
1 + \frac{\ln \Big( \tau^{el} \mbox{max}\{T, |\epsilon|\} \Big) }
{4\pi^2 g }
\right]  
\partial_{\tau}F_{\epsilon}(\tau) +
                                                        \label{k30}\\
&&\frac{ \partial_{\epsilon}F_{\epsilon}^{eq} }
{4\pi g} 
\int \!\! 
\frac {d \epsilon'}{2\pi}
\ln \Big( \tau^{el}   |\epsilon-\epsilon'| \Big)  
\partial_\tau F_{\epsilon'}(\tau) \, .
                                                        \nonumber
\end{eqnarray}  
We focus first on the logarithmic 
renormalization of the coefficient in front of $\partial_\tau F$. This 
coefficient corresponds to the charge $Z$ in Finkel'stein's terminology 
\cite{Finkelstein83}.  Eq.~(\ref{k30}) describes then the
renormalization of $Z$ (with the correct coefficient). We stress,
however, that in our  
theory renormalization of $Z$ takes place at the level of the saddle 
point equation for the effective action, and not as a result of the 
fluctuation corrections. This distinguishes $Z$ from the conductance,  
$g$, whose renormalization occurs only at the 
level of the one loop correction, see section \ref{subsec.AA} and 
Eq.~(\ref{l17}). Physically renormalization of $Z$ originates from 
the suppression of the single--particle DOS by the residual short range 
interactions. This effect is due to the fact that  single particle 
Hartree--Fock energies are shifted by the interactions in a way to reduce 
the DOS near the Fermi energy. We consider it very satisfactory that 
such purely mean--field effect is taken into account by the 
saddle point equation and not by  fluctuation corrections. 
The important thing, however, is to keep the last term of the expression
(\ref{k30}) as well. 
This is to say that only the ``out''--minus--``in'' combination 
has the physical meaning. Being considered together, as an integral 
operator acting on $F_{\epsilon'}$, these terms do not lead to divergent 
corrections.

\section{Discussion}
\label{sec.discussion}

We have developed a field theory for interacting disordered metals   
using the Keldysh dynamic formulation. The advantages of this
technique are twofold: (i) One avoids introduction of the replica
trick;  (ii) One naturally gains the ability to deal with
non--equilibrium situations. 
The latter manifests itself in the presence of the non--trivial object   
$F_{t,t'}(r)$, which plays the role of the fermionic distribution
function. The saddle point equation of the theory turns out to be the
kinetic equation which determines this function. No such object
is apparent in the replicated Matsubara formulation
\cite{Finkelstein83,Belitz94},
since by  construction it is limited to the equilibrium case. Based on
the analogy with spin glasses \cite{Sompolinsky}, one may speculate
that non--trivial solutions $F_{t,t'}(r)$ of the saddle point 
equation are analogous to the replica symmetry breaking solutions
of the saddle point equation in the replica formulation.

We mainly focussed our attention on the careful analysis on the saddle
equations of the theory.  In particular we suggested the
following two--step procedure: 

i) In the first step we account for the
purely phase effects of the fluctuating electric fields on the single
particle Green function by an appropriately chosen gauge transformation.
This enables us to get rid of the temporal variations
of the Green function which are not related to the quasiparticle
dynamics. The remaining temporal fluctuations of the Green functions
are associated with the particle dynamics and can be described in
terms of the quasiparticle  distribution function $F_{t,t'}(r)$.
This formulation ensures explicit gauge invariance of the kinetic
equation and preserves the continuity relations at every 
stage. As a byproduct of this procedure we were able to obtain a
non-perturbative expression for the DOS  --
the case where the phase effects give the main contribution. 
Such phase effects do not
contribute to gauge invariant observables which are
represented diagrammatically by closed loops. 
In the usual diagram technique this corresponds to a cancellation
between certain diagrams (the diagrams containing double logarithms in
two dimensions). By explicitly accounting for the phase effects we get
rid of these diagrams which significantly reduces the number of terms
in each order of perturbative expansion.

ii) After the phase effects have been taken into account, we obtain
a theory formulated in terms of the $\tilde Q$ matrix field.
The latter describes quantum  fluctuations of the electron distribution 
function in the close vicinity of the Fermi energy. Restricting
ourselves to the manifold of the massless fluctuations  
given by $\tilde Q^2=1$, we obtain the effective $\sigma$--model 
action, $S_{eff}[\tilde Q]$,  Eq.~(\ref{k2}). Searching then for the
extremum  of this action, we arrive at the kinetic equation on the
distribution  function. After this two--step saddle point procedure
one should consider  the quantum fluctuations effects. The
Altshuler--Aronov  corrections \cite{Altshuler85} to the conductance, $g$, 
turn out to be a manifestation of the one--loop quantum fluctuations. 

Although we have obtained renormalization of both parameters $g$ and $Z$, 
Eqs.~(\ref{l17}) and (\ref{k30}), we deliberately avoided putting it in the 
framework of the renormalization group.  The point is that after introducing 
the phase transformation and integrating out the photon fields, the effective 
action on $\tilde Q$, Eq.~(\ref{k2}), obtains a complicated form. We can not 
prove that this entire form is reproducible after the fast mode elimination.  
A seemingly better possibility is to perform the renormalization of
the action, which contains both $\tilde Q$ and $\nabla K$ fields,
Eq.~(\ref{q8}). In this case one has to specify how the relation
between $K$ and $\Phi$ fields changes in the process of
renormalization. Since we believe that the introduction of the phase,  
$K$, is a vital element of the theory, the more complicated form of
the action (compared to the one of Finkel'stein) is justifiable.  

We hope that the present formulation will help to shied light on 
the nature of the low--temperature phase of low--dimensional 
disordered metals. A few aspects of this theory seem to us very 
suggestive in this respect. Certain parallels with the spin 
glasses theory may prove to be useful. 
Apart from the extremely complicated problems  relating to the
character of the low--temperature phase, the functional Keldysh
formalism may be  useful for the description of non--equilibrium
effects in disordered metals and superconductors. 
The extension of this formalism to include the spin and Cooper
channels will be a subject of our future work.

\section{Acknowledgments}

We acknowledge the useful discussions with I. Aleiner, A. Altland,
B. Altshuler,  
Y. Gefen, D. Khmelnitskii, I. Lerner, B. Simons, I. Smolarenko 
and B. Spivak.  This research was supported by the NSF Grant No.~PHY
94-07194. The work was performed in part during the 1998 Extended Workshop
on Chaos and Interactions in Mesoscopic Physics in Trieste. It is our
pleasure  to thank the ICTP and the organizers of the workshop for their
hospitality.

\widetext

\begin{thebibliography}{34}
\bibitem{foot1} The independence of the normalization constant ${\cal N}$ from the
Hamiltonian follows from the fact that the operators entering the
Hamiltonian are normal ordered \cite{Negele}.
\bibitem{foot2} The vector source 
${\bf A}(r,t)$ that we are using here enters the electron Hamiltonian as 
$\hat H= (-i\nabla +{\bf A})^2/2m$ and therefore differs from the
vector potential by a factor of $e/c$. We shall refer to it however
as the vector potential.
\bibitem{foot4} Due to Eq.~(\ref{u2}) differentiation of $Z$ and 
$\ln Z$ is equivalent.
\end{thebibliography}
\end{document}